\documentclass[twocolumn]{aastex631}

\shorttitle{Joint JWST$+$ALMA imaging of $z\sim3$ SMGs }
\shortauthors{Hodge et al.}
\graphicspath{{./}{figures/}}

\begin{document}

\title{ALESS-JWST: Joint (sub-)kiloparsec JWST and ALMA imaging of $z\sim3$ submillimeter galaxies reveals heavily obscured bulge formation events}

\author[0000-0001-6586-8845]{J. A. Hodge}
\affiliation{Leiden Observatory, Leiden University, P.O. Box 9513, 2300 RA Leiden, the Netherlands}
\altaffiliation{hodge@strw.leidenuniv.nl}

\author[0000-0001-9759-4797]{E. da Cunha} 
\affiliation{International Centre for Radio Astronomy Research, University of Western Australia, 35 Stirling Hwy, Crawley 26WA 6009, Australia}
\affiliation{ARC Centre of Excellence for All Sky Astrophysics in 3 Dimensions (ASTRO 3D), Australia}

\author[0000-0002-7612-0469]{S. Kendrew} 
\affiliation{European Space Agency (ESA), ESA Office, Space Telescope Science Institute, 3700 San Martin Drive, Baltimore, MD 21218, USA}

\author[0000-0002-8184-5229]{J. Li}
\affiliation{International Centre for Radio Astronomy Research, University of Western Australia, 35 Stirling Hwy, Crawley 26WA 6009, Australia}

\author[0000-0003-3037-257X]{I. Smail}
\affiliation{Centre for Extragalactic Astronomy, Department of Physics, Durham University, South Road, Durham DH1 3LE, UK}

\author[0009-0004-3732-6394]{B. A. Westoby}
\affiliation{Leiden Observatory, Leiden University, P.O. Box 9513, 2300 RA Leiden, the Netherlands}

\author[0000-0001-6576-6339]{O. Nayak}
\affiliation{European Space Agency (ESA), ESA Office, Space Telescope Science Institute, 3700 San Martin Drive, Baltimore, MD 21218, USA}
\affiliation{NASA Goddard Space Flight Center, 8800 Greenbelt Road, Greenbelt, MD, USA}

\author[0000-0003-1192-5837]{A. M. Swinbank}
\affiliation{Centre for Extragalactic Astronomy, Department of Physics, Durham University, South Road, Durham DH1 3LE, UK}

\author[0000-0002-3805-0789]{C.-C. Chen}
\affiliation{Academia Sinica Institute of Astronomy and Astrophysics (ASIAA), No. 1, Sec. 4, Roosevelt Road, Taipei 106216, Taiwan}

\author[0000-0003-4793-7880]{F. Walter}
\affiliation{Max--Planck Institut f\"ur Astronomie, K\"onigstuhl 17, 69117 Heidelberg, Germany}

\author[0000-0001-5434-5942]{P. van der Werf}
\affiliation{Leiden Observatory, Leiden University, P.O. Box 9513, 2300 RA Leiden, the Netherlands}

\author[0000-0002-7698-3002]{M. Cracraft}
\affiliation{Space Telescope Science Institute, 3700 San Martin Drive, Baltimore, MD 21218, USA}

\author[0000-0003-4569-2285]{A. Battisti}
\affiliation{Research School of Astronomy and Astrophysics, Australian National University, Cotter Road, Weston Creek, ACT 2611, Australia}
\affiliation{International Centre for Radio Astronomy Research, University of Western Australia, 35 Stirling Hwy, Crawley 26WA 6009, Australia}
\affiliation{ARC Centre of Excellence for All Sky Astrophysics in 3 Dimensions (ASTRO 3D), Australia}

\author[0000-0002-0167-2453]{W. N. Brandt}
\affiliation{Department of Astronomy and Astrophysics, 525 Davey Lab, The Pennsylvania State University, University Park, PA 16802, USA}
\affiliation{Institute for Gravitation and the Cosmos, The Pennsylvania State University, University Park, PA 16802, USA}
\affiliation{Department of Physics, 104 Davey Laboratory, The Pennsylvania State University, University Park, PA 16802, USA}

\author[0000-0003-0085-6346]{G. Calistro Rivera}
\affiliation{German Aerospace Center (DLR), Institute of Communications and Navigation, Wessling, Germany}
\affiliation{European Southern Observatory (ESO), Karl-Schwarzschild-Straße 2, 85748 Garching bei M¨unchen, Germany}

\author[0000-0002-8487-3153]{S. C. Chapman}
\affiliation{Department of Physics and Atmospheric Science, Dalhousie University, 6310 Coburg Road, B3H 4R2, Halifax, Canada}
\affiliation{National Research Council, Herzberg Astronomy and Astrophysics, 5071 West Saanich Road, Victoria, V9E 2E7, Canada}
\affiliation{Department of Physics and Astronomy, University of British Columbia, 6225 Agricultural Road, Vancouver, V6T 1Z1, Canada}
\affiliation{Eureka Scientific Inc, Oakland, CA 94602, USA}

\author[0000-0003-2027-8221]{P. Cox}
\affiliation{Sorbonne Universit{\'e}, UPMC Universit{\'e} Paris 6 and CNRS, UMR 7095, Institut d'Astrophysique de Paris, 98bis boulevard Arago, 75014 Paris, France}

\author[0000-0001-7147-3575]{H. Dannerbauer}
\affiliation{Instituto de Astrof\'isica de Canarias, V\'ia L\'actea, 39020 La Laguna (Tenerife), Spain}

\author[0000-0002-2662-8803]{R. Decarli}
\affiliation{INAF -- Osservatorio di Astrofisica e Scienza dello Spazio di Bologna, via Gobetti 93/3, I-40129, Bologna, Italy}

\author[0000-0002-9278-7028]{M. Frias Castillo}
\affiliation{Leiden Observatory, Leiden University, P.O. Box 9513, 2300 RA Leiden, the Netherlands}

\author[0000-0002-2554-1837]{T. R. Greve}
\affiliation{Cosmic Dawn Center (DAWN), Denmark}
\affiliation{DTU Space, Technical University of Denmark, Elektrovej, Building 328, 2800, Kgs. Lyngby, Denmark}
\affiliation{Dept. of Physics and Astronomy, University College London, Gower Street, London WC1E 6BT, United Kingdom}

\author[0000-0002-7821-8873]{K. K. Knudsen}
\affiliation{Department of Space, Earth and Environment, Chalmers University of Technology, SE-412 96 Gothenburg, Sweden}

\author[0000-0002-4826-8642]{S. Leslie}
\affiliation{Leiden Observatory, Leiden University, P.O. Box 9513, 2300 RA Leiden, the Netherlands}

\author[0000-0001-6459-0669]{K. M. Menten}
\affiliation{Max-Planck-Institut f\"ur Radioastronomie, Auf dem H\"ugel 69, D-53121 Bonn, Germany}

\author[0000-0002-1383-0746]{M. Rybak}
\affiliation{Leiden Observatory, Leiden University, P.O. Box 9513, 2300 RA Leiden, the Netherlands}
\affiliation{Faculty of Electrical Engineering, Mathematics and Computer Science, Delft University of Technology, Mekelweg 4, 2628 CD Delft, The Netherlands}
\affiliation{SRON -- Netherlands Institute for Space Research, Niels Bohrweg~4, 2333 CA Leiden, The Netherlands}

\author[0000-0002-3933-7677]{E. Schinnerer}
\affiliation{Max--Planck Institut f\"ur Astronomie, K\"onigstuhl 17, 69117 Heidelberg, Germany}

\author[0000-0003-2376-8971]{J. L. Wardlow}
\affiliation{Department of Physics, Lancaster University, Lancaster, LA1 4YB, UK}

\author[0000-0003-4678-3939]{A. Weiss}
\affiliation{Max-Planck-Institut f\"ur Radioastronomie, Auf dem H\"ugel 69, D-53121 Bonn, Germany}



\begin{abstract}

We present JWST NIRCam imaging targeting 13 $z$\,$\sim$\,3 infrared-luminous ($L_{\rm IR}\sim5\times10^{12}L_{\odot}$) galaxies from the ALESS survey with uniquely deep, high-resolution (0.08$''$--0.16$''$) ALMA 870$\mu$m imaging. The 2.0-4.4$\mu$m (observed frame) NIRCam imaging reveals the rest-frame near-infrared stellar emission in these submillimeter-selected galaxies (SMGs) at the same (sub-)kpc resolution as the 870$\mu$m dust continuum. The newly revealed stellar morphologies show striking similarities with the dust continuum morphologies at 870$\mu$m, with the centers and position angles agreeing for most sources, clearly illustrating that the spatial offsets reported previously between the 870$\mu$m and HST morphologies were due to strong differential dust obscuration. The F444W sizes are 78$\pm$21\% larger than those measured at 870$\mu$m, in contrast to recent results from hydrodynamical simulations that predict larger 870$\mu$m sizes. We report evidence for significant dust obscuration in F444W for the highest-redshift sources, emphasizing the importance of longer-wavelength MIRI imaging. The majority of the sources show evidence that they are undergoing mergers/interactions, including tidal tails/plumes---some of which are also detected at 870$\mu$m. We find a clear correlation between NIRCam colors and 870$\mu$m surface brightness on $\sim$1\,kpc scales, indicating that the galaxies are primarily red due to dust---not stellar age---and we show that the dust structure on $\sim$kpc-scales is broadly similar to that in nearby galaxies. Finally, we find no strong stellar bars in the rest-frame near-infrared, suggesting the extended bar-like features seen at 870$\mu$m are highly obscured and/or gas-dominated structures that are likely early precursors to significant bulge growth. 

\end{abstract}



\section{Introduction} \label{sec:Introduction}

With the advent of the first infrared (IR) sky surveys, including by IRAS \citep{Neugebauer_1984} and COBE \citep{Mather_1994COBE},  
a basic fact of the Universe was revealed: Roughly half of the star formation that has occurred over cosmic time took place shrouded behind cosmic dust \citep[e.g.,][]{Puget_1996, Hauser_1998}. 
Thanks to the decades of dedicated follow-up that subsequently occurred in order to systematically resolve the IR/sub-millimeter background, 
we now know that this dust-obscured star formation was as important as unobscured star formation at least out to $z\sim4$ \cite[e.g.,][]{Dudzeviciute_2020, Bouwens_2020, Zavala_2021}, and perhaps even beyond \citep[e.g.,][]{Algera_2023}. 
The galaxies with the 
most significant dust-obscured star formation rates are the $z\sim2-5$ galaxies discovered in the earliest single-dish submillimeter surveys \citep[`SMGs'; e.g.,][]{Smail_1997, Hughes_1998, Barger_1998, Eales_1999}. Canonically, SMGs are thought to be tied to the evolution of QSOs \citep[e.g.,][]{sanders1996, Hopkins2006} and ultimately to the build-up of massive elliptical galaxies \citep[e.g.,][and see \citealt{Casey_2014} for a review]{Smail_2004, Swinbank2006, Cimatti2008, vanDokkum2008, Simpson_2014, Toft_2014}. Understanding the physical processes that drove their intense dusty star-formation is thus crucial for understanding massive galaxy evolution. 

Prior to the launch of the James Webb Space Telescope (JWST), characterizing the stellar counterparts of SMGs was notoriously difficult. Studies using, e.g., deep HST H-band imaging report conflicting results on the prevalence of merger activity \citep[e.g.,][]{Swinbank_2010, Targett_2013, Chen_2015}. However, these dust-rich
galaxies are particularly prone to structured dust extinction \citep[e.g.,][]{Simpson2017, Cochrane_2019}, 
suggesting their rest-frame UV/optical classifications may be unreliable \citep[e.g.,][]{Lang_2019, Popping_2022}. 
Moreover, such studies also report high fractions of `NIR-faint' sources \citep[$\sim$20\% at $m_{H}$$\sim$27.8 mag depth;][]{Chen_2015}, presumably due to their high levels of dust attenuation \citep[median global $A_{\rm V}$$\sim$2--3\,mag; ][]{daCunha_2015, Dudzeviciute_2020}, precluding reliable morphological classification. Finally, even for the sources detected with, e.g., HST, their stellar masses typically carry large uncertainties due to the well-known degeneracy between stellar age and dust attenuation   
\citep[e.g.,][]{Hainline_2011, Michalowski_2014, Dudzeviciute_2020}. 
These observational challenges have contributed to the lack of theoretical consensus on the nature of SMGs and the physical mechanism(s) by which their dusty star formation was triggered \citep[e.g.,][]{Baugh_2005, Narayanan_2010, Narayanan_2015,  Hayward_2013a, Hayward_2013b, McAlpine_2019, Lovell_2021}.

In the last decade, the advent of the Atacama Large Millimeter/submillimeter Array (ALMA) has provided critical new insight into dust-obscured star formation at high redshift (e.g., see \citealt{Hodge_2020} for a review). 
In particular, ALMA (along with the Submillimeter Array, SMA, and the Northern Extended Millimeter Array, NOEMA) have allowed large samples of the classically single-dish-selected SMGs to be located with interferometric accuracy  \citep[e.g.,][]{Younger_2008MNRAS.387..707Y, Younger_2009ApJ...704..803Y, Barger2012, Smolcic_2012, Hodge_2013, Simpson_2014, Spilker_2016, Hill_2018, Stach_2019} and resolved directly in their (sub-)millimeter continuum emission \citep[e.g.,][]{Simpson_2015_sizes, Ikarashi_2015, Hodge_2016, Gullberg_2019}. 
These studies have revealed that the (sub-)millimeter continuum emission in these galaxies is relatively compact and has disk-like morphology \citep[i.e., with a S\'ersic index $n\sim 1$; e.g., ][]{Simpson_2015_sizes, Hodge_2016, Gullberg_2019} and often appears spatially offset from the rest-frame UV/optical continuum \citep[e.g.,][]{Hodge_2012, Chen_2015, CalistroRivera_2018}, which some studies have interpreted as being due to real physical offsets between the location of the dusty star formation and the bulk of the stellar mass \citep{Chen_2015}. The highest-resolution studies have further resolved the (sub-)millimeter continuum-traced dust disks in small sub-samples into distinct substructures on $\sim$0.5--1\,kpc scales, uncovering features with morphologies suggestive of bars, rings, and (tidally-induced) spiral arms (\citealt{Hodge_2019}, and see also \citealt{Gullberg_2019}). Unfortunately, despite the depth of the available HST imaging, no meaningful comparison could be made to the stellar structure due to the faintness of the sources in the rest-frame UV/optical.

Recently, the launch of JWST has shed new light on the nature of submillimeter-selected sources by revealing their rest-frame near-infrared morphologies on (sub-)kpc scales \citep[e.g.,][]{Chen_2022, Cheng_2023, Colina_2023, Huang_2023, Gillman_2023, Gillman_2024, Boogaard_2024, CrespoGomez_2024}. 
Some of these early studies have reported evidence for undisturbed, disk-like morphologies that have been taken as evidence for the importance of secular growth for their stellar mass assembly \citep[e.g.,][]{Cheng_2023}. Other studies have also reported evidence for disk-like morphologies but with a broad scatter in the non-parametric parameter space for morphological classification  \citep{Gillman_2023}, indicating a diverse population that nevertheless reflects the field population \citep{Gillman_2024}. A number of studies have identified structures in the rest-frame near-infrared emission akin to bulges, spiral arms, and bars \citep[e.g.,][]{Chen_2022, Huang_2023, Smail_2023, Amvrosiadis_2024}, as well as examples of tidal features indicating dynamical interactions \citep[e.g.,][]{Chen_2022, Cathey_2024}. 
However, given the relatively narrow-field coverage of JWST in relation to the number density of SMGs, the submillimeter-selected galaxies studied so far in extragalactic deep fields have largely constituted submillimeter-fainter sources (i.e., $\lesssim$few mJy at 850$\mu$m), which may undergo different triggering mechanisms than the submillimeter-brighter sources \citep[e.g.,][]{Hayward_2013a}.  More crucially, the current samples have lacked equally high-resolution ALMA dust continuum imaging, preventing the incorporation of the (energetically important) dust continuum emission into the physical interpretation.

Here, we present multi-band NIRCam imaging at 2--4.4$\mu$m (observed-frame) of 13 $z$\,$\sim$\,3 SMGs from the ALMA follow-up of the LABOCA Extended \textit{Chandra} Deep Field South Survey \citep[ALESS;][]{Hodge_2013, Karim_2013}. 
These submillimeter-bright (median $S_{\rm 870}$\,$\sim$\,6\,mJy) galaxies all have deep ALMA 870$\mu$m continuum imaging at 0.08$''$--0.16$''$ resolution \citep{Hodge_2016, Hodge_2019}, 
allowing us to compare their newly revealed rest-frame near-infrared emission to their dust continuum emission on the same (sub-)kpc scales. We begin in \S\ref{sec:data} by discussing the sample selection, 
the JWST observations and data reduction, and the existing 870$\mu$m ALMA data. We present our analysis and results in \S\ref{sec:analysis}, including a comparison between the detected stellar and dust continuum emission, a visual classification, 
a curve-of-growth analysis of the effective radii, {\sc Galfit} modeling, isophotal fitting to search for stellar bars, and finally a comparison of the NIRCam colors with ALMA 870$\mu$m. We follow this with a discussion in \S\ref{sec:discussion}. We end with our conclusions in \S\ref{sec:Conclusions}. 
Throughout this paper, we adopt a standard spatially flat $\Lambda$CDM cosmology from the Planck 2018 results with Hubble constant $H_{0}$ $=$ 67.4\,km s$^{-1}$ Mpc$^{-1}$, dark energy density parameter $\Omega_{\Lambda}$ $=$ 0.685, and matter density parameter $\Omega_{M}$ $=$ 0.315 \citep{Planck2020}. With these parameters, 1$''$ corresponds to 7.9\,kpc at $z\sim3$.
We adopt a \citet{Chabrier_2003} initial mass function (IMF) and AB system magnitudes.

\def\baselinestretch{1.1} 

\begin{deluxetable*}{lcccccc}
	\tablecolumns{6}
	\tabletypesize{\footnotesize}
	\tablecaption{Target Properties: Sample}
	\tablehead{
        \colhead{Source ID\tablenotemark{a}} &
        \colhead{Position\tablenotemark{a}} &
        \colhead{$S_{\rm 870}$\tablenotemark{a}} &
	\colhead{$z$\tablenotemark{b}} &
        \colhead{Redshift source\tablenotemark{b}} &
        \colhead{870$\mu$m} &
        \colhead{NIRCam}\\
        &
        (J2000) &
        (mJy) &
        &
        &       
        FWHM\tablenotemark{c} & 
        pointing
	}
	\startdata
ALESS 1.1  &  03:33:14.46 $-$27:56:14.5  & 6.7 $\pm$ 0.5 & 4.674 & CO(5--4)  & 0.16$''$ & 2 \\
ALESS 1.2  &  03:33:14.41 $-$27:56:11.6  & 3.5 $\pm$ 0.4 & 4.669  & CO(5--4)  & 0.16$''$ & 2 \\
ALESS 1.3  &  03:33:14.16 $-$27:56:12.5  & 1.9 $\pm$ 0.4 & 2.86$^{+0.46}_{-1.53}$   & $z_{\textnormal{phot}}$  & 0.16$''$ & 2 \\
ALESS 3.1  &  03:33:21.51 $-$27:55:20.5  & 8.3 $\pm$ 0.4 & 3.375  & CO(5--4), CO(4--3), [CI]  & 0.08$''$ & 2\\
ALESS 9.1  &  03:32:11.33 $-$27:52:12.0  & 8.8 $\pm$ 0.5 & 3.694  & CO(4--3), [CI]  & 0.08$''$ & 3 \\
ALESS 10.1 &  03:32:19.05 $-$27:52:14.8  & 5.2 $\pm$ 0.5 & 3.34$^{+0.02}_{-0.16}$   & $z_{\textnormal{phot}}$\tablenotemark{d}  & 0.16$''$ & 3 \\
ALESS 15.1  &  03:33:33.37 $-$27:59:29.7  & 9.0 $\pm$ 0.4 & 2.86$^{+0.10}_{-0.20}$  & $z_{\textnormal{phot}}$  & 0.08$''$ & 1 \\
ALESS 17.1 &  03:32:07.29 $-$27:51:20.9  & 8.4 $\pm$ 0.5 & 1.5397 & H$_{\alpha}$, CO(5--4), CO(2--1)  & 0.08$''$ & 3 \\
ALESS 29.1  &  03:33:36.90 $-$27:58:09.3  & 5.9 $\pm$ 0.4 & 3.69$^{+0.42}_{-0.45}$ & $z_{\textnormal{phot}}$\tablenotemark{e}  & 0.16$''$ & 1 \\
ALESS 45.1  &  03:32:25.26 $-$27:52:30.6  & 6.0 $\pm$ 0.5 & 3.09$^{+0.22}_{-0.26}$  & $z_{\textnormal{phot}}$  & 0.16$''$ & 3 \\
ALESS 76.1 &  03:33:32.35 $-$27:59:55.7  & 6.4 $\pm$ 0.6 & 3.3895  & [OIII]  & 0.08$''$ & 1 \\
ALESS 112.1  &  03:32:48.86 $-$27:31:13.2  & 7.6 $\pm$ 0.5 & 2.314  & Ly$\alpha$, CO(3--2)  & 0.08$''$ & 4 \\
\hline
ALESS 3.1-comp & 03:33:21.43 $-$27:55:25.4 & 1.07 $\pm$ 0.06 & 3.374\tablenotemark{f} & CO(5--4), CO(4--3)\tablenotemark{f} & 0.08$''$ & 2\
    \enddata
    \tablenotetext{a}{Source IDs and 870$\mu$m flux densities are from \citet{Hodge_2013} except for ALESS 3.1-comp, which is newly reported here. Source positions for all sources have been updated based on the high-resolution 870$\mu$m data available in \citet{Hodge_2016, Hodge_2019}.}
    \tablenotetext{b}{Rest-frame optical/UV-based spectroscopic redshifts are from \citet{Danielson_2017}, CO-based redshifts are from \citet{Birkin_2021}, and the photometric redshifts were taken from \citet{daCunha_2015} and updated to include the new NIRCam$+$MIRI photometry (Li et al. in prep.).}
    \tablenotetext{c}{Highest-resolution 870$\mu$m continuum data available on the source. All sources with 0.08$''$ observations also have 0.16$''$ observations. For details, see Section~\ref{sec:ALMAdata}.} 
    \tablenotetext{d}{ALESS 10.1 was previously reported to have $z_{\rm spec}$ $=$ 0.7616 based on multiple lines in an optical spectrum \citep{Danielson_2017}, but this was likely an incorrect identification of an unrelated foreground source based on the very red colors of the SMG, which are difficult to reconcile with such a redshift. We therefore assume its photometric redshift here.}
    \tablenotetext{e}{ALESS 29.1 was previously reported to have $z_{\rm spec}$ $=$ 1.438 \citep{Danielson_2017}, but based on both its red colors (difficult to reconcile with this redshift) and the lack of a CO detection at that redshift \citep{Birkin_2021}, we assume its photometric redshift here.}
    \tablenotetext{f}{CO-based redshift from Westoby et al.\,(in prep.)}
	\label{tab:sample}
\end{deluxetable*}
\def\baselinestretch{1.0} 

\section{Observations \& data reduction}
\label{sec:data}

\subsection{Sample Selection}
\label{sec:sample}

The JWST observations targeted 13 SMGs\footnote{The 13th SMG, ALESS 3.1-comp, was discovered by the subsequent deep 870$\mu$m imaging \citep{Hodge_2016, Hodge_2019} near ALESS 3.1; see Section~\ref{sec:JWSTvsALMA} for details.} originally detected in the LABOCA 870$\mu$m survey of the ECDFS \citep{Weiss_2009} and then interferometrically identified as part of the ALESS survey \citep{Hodge_2013, Karim_2013, Simpson_2014, daCunha_2015, Danielson_2017}. These 13 galaxies were part of the sample subsequently followed up with high-resolution, high-fidelity ALMA 870$\mu$m continuum mapping observations \citep[][and see Section~\ref{sec:ALMAdata} for details]{Hodge_2016, Hodge_2019}. 
The sample is listed in Table~\ref{tab:sample}. 
Eight of the targets have spectroscopic redshifts, also listed in Table~\ref{tab:sample}; the remaining five have high-quality {\sc Magphys}-based photometric redshifts which have been updated from those listed in \citet{daCunha_2015} to include the new JWST NIRCam$+$MIRI photometry (Li et al. in prep; Table~\ref{tab:sample}).

Due to the original selection criteria for the high-resolution ALMA 870$\mu$m follow-up, which was based on total 870$\mu$m flux density and the availability of (effectively randomly targeted) HST coverage \citep{Chen_2015}, the sources targeted here include some of the submillimeter-brightest sources from the ALESS parent sample: the median 870$\mu$m flux density of the sources is 6.4\,mJy, with a 16th -- 84th percentile range of 3.4--8.4\,mJy. The galaxies nevertheless span a wide range of redshifts ($z\simeq 1.5-4.5$) and star formation rates ($\sim$60--1000 M$_{\odot}$ yr$^{-1}$).  
The median redshift of the targets is 3.4, with a 16th -- 84th percentile range of 2.8 to 3.8. 
The median SFR is 350 M$_{\odot}$ yr$^{-1}$, with a 16th -- 84th percentile range of 190--820 M$_{\odot}$ yr$^{-1}$. The sample is thus slightly higher-redshift and more highly star-forming than the parent population of ALESS SMGs \citep[$z_{\rm phot}$ 
$=$ 2.7 $\pm$ 0.1 and SFR $=$ 280 $\pm$ 70 M$_{\odot}$ yr$^{-1}$;][]{daCunha_2015}. Finally, two of the sources (ALESS 17.1 and 45.1) are associated with known X-ray sources, of which one (ALESS 17.1) was identified as an X-ray AGN by \citet{Wang_2013}, while they found the X-ray source in ALESS 45.1 could be plausibly explained by star formation. The presence of a bright AGN could enhance the central brightness of the observed stellar distribution and lead to a higher S\'ersic index; 
we note that this will not affect the majority of the sample. We also note that the X-ray source in ALESS 17.1 appears to be associated with a companion to the 870$\mu$m-bright SMG (see Fig.~\ref{fig:JWSTgalfit}).

\subsection{JWST Observations \& Data Reduction}
\label{sec:JWSTobs}

The JWST data presented here are part of a GO Cycle~1 Program (PID 2516, PIs: Hodge \& da Cunha) to perform near- and mid-infrared imaging of submillimeter galaxies from the~\citet{Hodge_2016, Hodge_2019} samples, as described above. The 13 primary targets were selected from the \citet{Hodge_2016, Hodge_2019} samples such that they could be covered with four pointings with the NIRCam instrument. Each field is imaged in the F200W, F356W and F444W filters, using the `INTRAMODULEBOX' 4-point dither pattern in order to cover a compact square region without gaps. 
Three NIRCam pointings used Module B only, the 4$^{\rm th}$ used both Modules A and B; with an on-source exposure time of 30 mins per field. The observations were executed between 2022 September and December. One observation was heavily impacted by slew artifacts on first execution, having been executed immediately following NIRCam observations of a very bright target; this was repeated in 2022 November.

The data were reduced and calibrated using version 1.11.3 of the \texttt{jwst} calibration pipeline \citep{bushouse_howard_2023_7829329} and context 1119 of the JWST Calibration Reference Data System (CRDS). We used the standard JWST calibration pipeline, broadly following the recipe and modifications from the CEERS survey~\citep{Bagley2023}. The first pipeline stage, performing detector calibrations, was run with snowball detections enabled and a jump step detection threshold of 6$\sigma$. The $1/f$ noise striping pattern was removed using the \texttt{remstriping} algorithm\footnote{https://github.com/ceers/ceers-nircam}, before proceeding with the standard settings for the \texttt{calwebb\_image2} pipeline, which returns photometrically calibrated images for each exposure. The third pipeline stage (\texttt{calwebb\_image3}) was run in stages with some modifications, following~\citet{Bagley2023}; this includes the background subtraction, which uses a tiered source mask approach to compute and subtract a pedestal value for the sky, and adjusts the variance arrays in the background-subtracted images accordingly. The resulting photometry for the targets is consistent with expectations based on the MUSYC catalog \citep{Cardamone_2010} 
and will be discussed further in Li et al. (in prep.). The PSF FWHM in the F200W, F356W, and F444W filters is 0.07$''$, 0.12$''$, and 0.14$''$, respectively. 
For the redshift range [median redshift] of our targets, the F200W, F356W, and F444W filters will trace rest-frame 0.35--0.8$\mu$m [0.5$\mu$m], 
0.6--1.4$\mu$m [0.8$\mu$m],
and 0.8--1.8$\mu$m [1$\mu$m] stellar continuum emission\footnote{We note that the H$\alpha$ and [OIII]$\lambda\lambda$ 5007,4959 emission lines may fall in the F200W or F356W filters for the redshift range of our targets. However, given the wide filter bandwidths and relatively low equivalent width lines expected for these highly dust-obscured sources, this is not expected to impact the current analysis.}, respectively, with the F444W resolution corresponding to physical scales of $\sim$0.9$-$1.2\,kpc for our range of target redshifts. A common pixel sampling of 0.03$''$ was used for all filters. 

Given the requirement for precise alignment to the ALMA data, the astrometric calibration is an important aspect of the data reduction, and this calibration step proved challenging due to the lack of point sources in our fields. The \texttt{tweakreg} step was used to perform both the relative and absolute alignments, the latter using the Gaia DR3 catalog~\citep{GaiaDR3}. However, some modifications were used for the individual pointings to optimize the outcome of the adjustment. These include removing Gaia sources without proper motion information from the reference catalog\footnote{See https://github.com/spacetelescope/jwst/issues/8168 for a discussion of this issue.}; and removing poorly-centroided sources from the pipeline-created source catalogs and/or using PSF photometry to improve centroids, especially for partially-saturated sources. 

The resulting relative and absolute astrometric accuracy was estimated from the final mosaics, by comparing measured point-source centroids between filters, and computing the average offset between the NIRCam and Gaia DR3 catalog coordinates, respectively. The resulting astrometric accuracies are shown in 
the Appendix. We also list 
the number of Gaia stars available for absolute alignment. We estimate a median absolute uncertainty in the astrometric solutions of 0.03$''$ (i.e., one pixel). 

\begin{figure*}
\centering
\includegraphics[width=\textwidth]{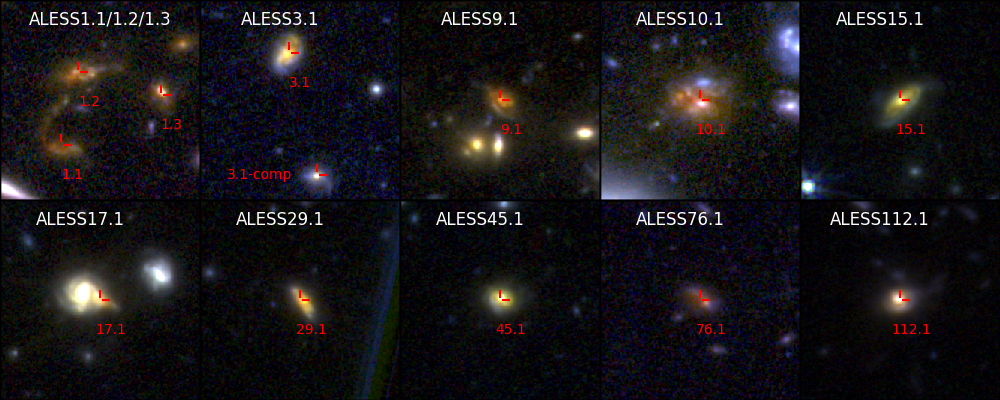}
\caption{JWST NIRCam RGB images of 8$''$$\times$8$''$ regions ($\sim$65$\times$65\,kpc at $z\sim3$) around FIR-luminous SMGs from the ALMA ALESS survey. North is up and East is to the left. The RGB images were made with the F444W (red), F356W (green), and F200W (blue) NIRCam filters. The positions of the 870$\mu$m sources are indicated with their ALESS ID numbers. All galaxies are clearly detected by NIRCam. ALESS 1.1, 1.2, and 1.3 are shown together in the same cutout. Note the long tidal tail revealed by the NIRCam imaging of ALESS 1.1/1.2, indicating an interaction. There is also an 870$\mu$m-detected companion (3.1-comp) to the South of ALESS 3.1 with the same spectroscopic redshift (see Section~\ref{sec:JWSTvsALMA} for details).}
\label{fig:RGBfields}
\end{figure*}

\begin{figure*}
\centering
\includegraphics[trim={1.2cm 1cm 1.2cm 0.5cm},clip, width=0.49\textwidth]{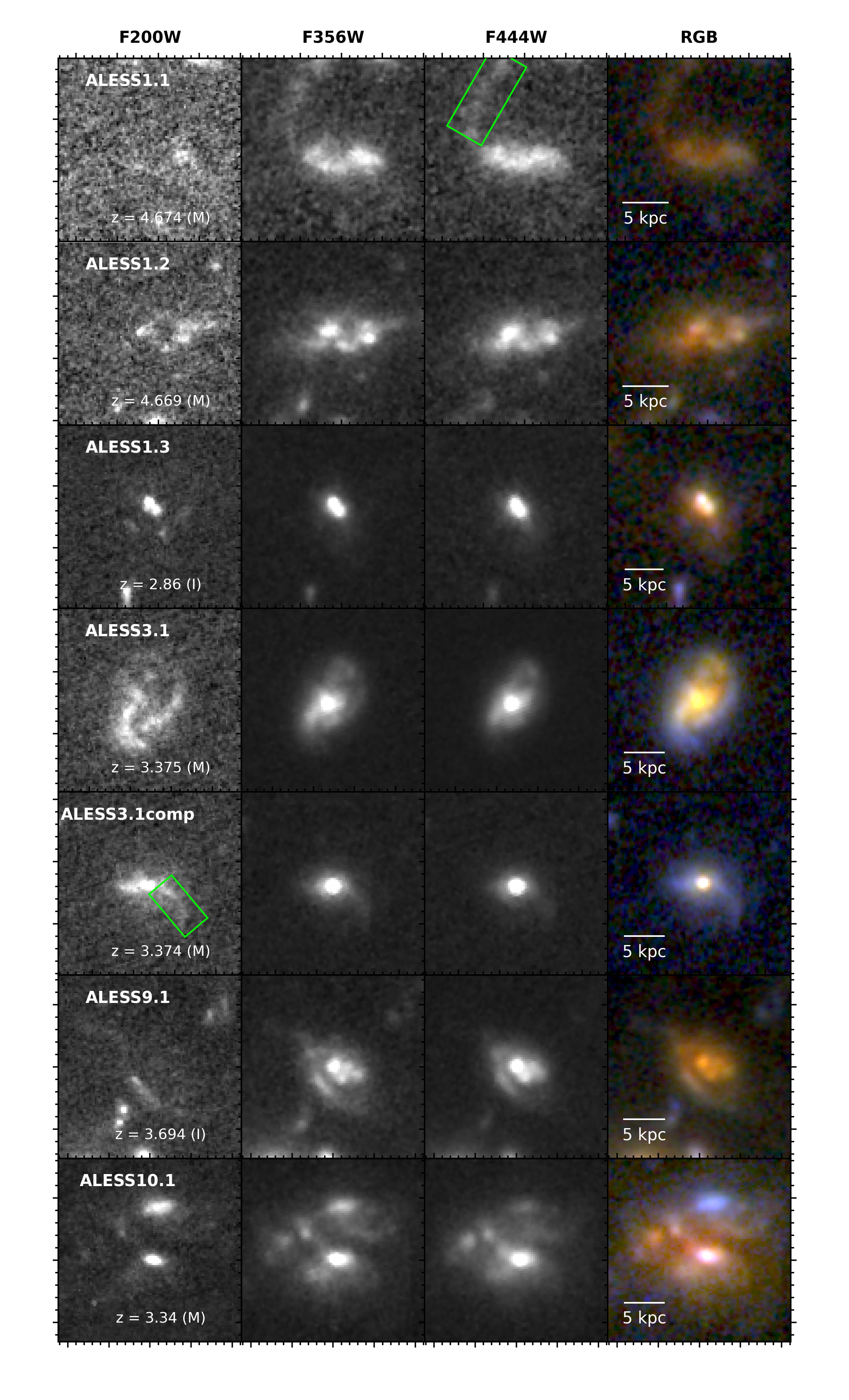}
\hfill
\includegraphics[trim={1.2cm 1cm 1.2cm 0.5cm},clip, width=0.49\textwidth]{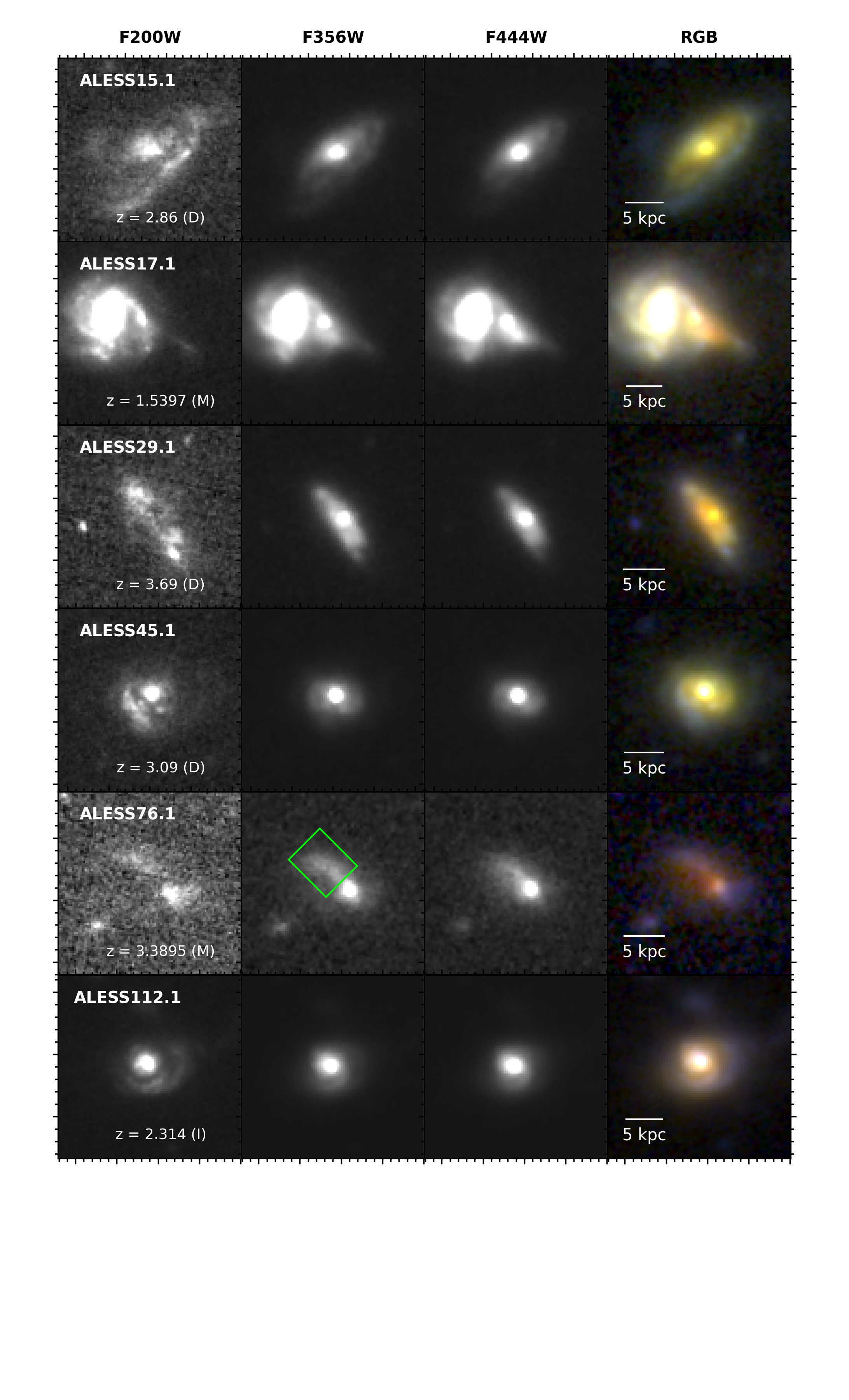}
\caption{JWST NIRCam cutouts (3$''$$\times$3$''$, or $\sim$25$\times$25\,kpc at $z\sim3$)  
of the 13 sources. For each galaxy, the columns show (from left): F200W, F356W, F444W, and the corresponding RGB image. Individual filters are scaled with a linear stretch between 0.25\% and 99.5\%, except for ALESS 17.1, where a maximum of 95.0\% is used to highlight the faint red submillimeter galaxy to the West and the brighter companion to the East. Most of the targeted SMGs are heavily dust-attenuated even with NIRCam's 2$\mu$m filter but are significantly less obscured above 3.5$\mu$m ($\simeq$1$\mu$m rest-frame). Some examples of apparent tidal tails are highlighted with boxes in the filter where they are most visible, and the panels are labeled with redshift and visual classification (Disk, Indeterminate, or Merger; See Section~\ref{sec:classification}.) }
\label{fig:NRCfilters}
\end{figure*}

\subsection{ALMA Data}
\label{sec:ALMAdata}
The ALMA continuum observations utilized for the primary targets in this study were taken as part of programs 2012.1.00307.S and 2016.1.00048.S and previously presented in \citet{Hodge_2016, Hodge_2019}, respectively. For both studies, the frequency setup was centered at 344\,GHz (Band 7; 870$\mu$m observed frame) with 4$\times$128 dual polarization channels covering the 8\,GHz bandwidth in Time Division Mode (TDM). At the median redshift of our targets ($z\sim3$), 870$\mu$m corresponds to 210$\mu$m in the rest-frame. 
The naturally weighted continuum data were imaged with pixel scales of 0.02$''$ and 0.01$''$ and achieved angular resolutions of 0.17$''$$\times$0.15$''$ and 0.10$''$$\times$0.07$''$ for the analyses of \citet{Hodge_2016} and \citet{Hodge_2019}, respectively (corresponding to physical scales of $\sim$1.3$\times$1.2\,kpc and $\sim$800$\times$550\,pc at a redshift of $z\sim3$). This range of spatial resolutions is similar to that achieved by NIRCam from 2.0-4.4$\mu$m (0.07$\mathrm{\arcsec}$-0.14$\mathrm{\arcsec}$). The typical RMS noise achieved in the 870$\mu$m continuum images is 64\,$\mu$Jy beam$^{-1}$ (for the 0.16$''$ images) and 20\,$\mu$Jy beam$^{-1}$ (for the 0.08$''$ images). 
Given the angular resolution of the 870$\mu$m observations and the high SNR of the targets, the \textit{statistical} astrometric accuracy 
is likely limited by the phase variations over the
array to a few [10] mas for images with 0.08$''$ [0.16$''$] resolution\footnote{ALMA Cycle 11 Technical Handbook, Chapter 10.5.2}.

\begin{figure*}
\centering
\includegraphics[trim={1cm 1cm 1cm 1cm},clip, width=0.49\textwidth]{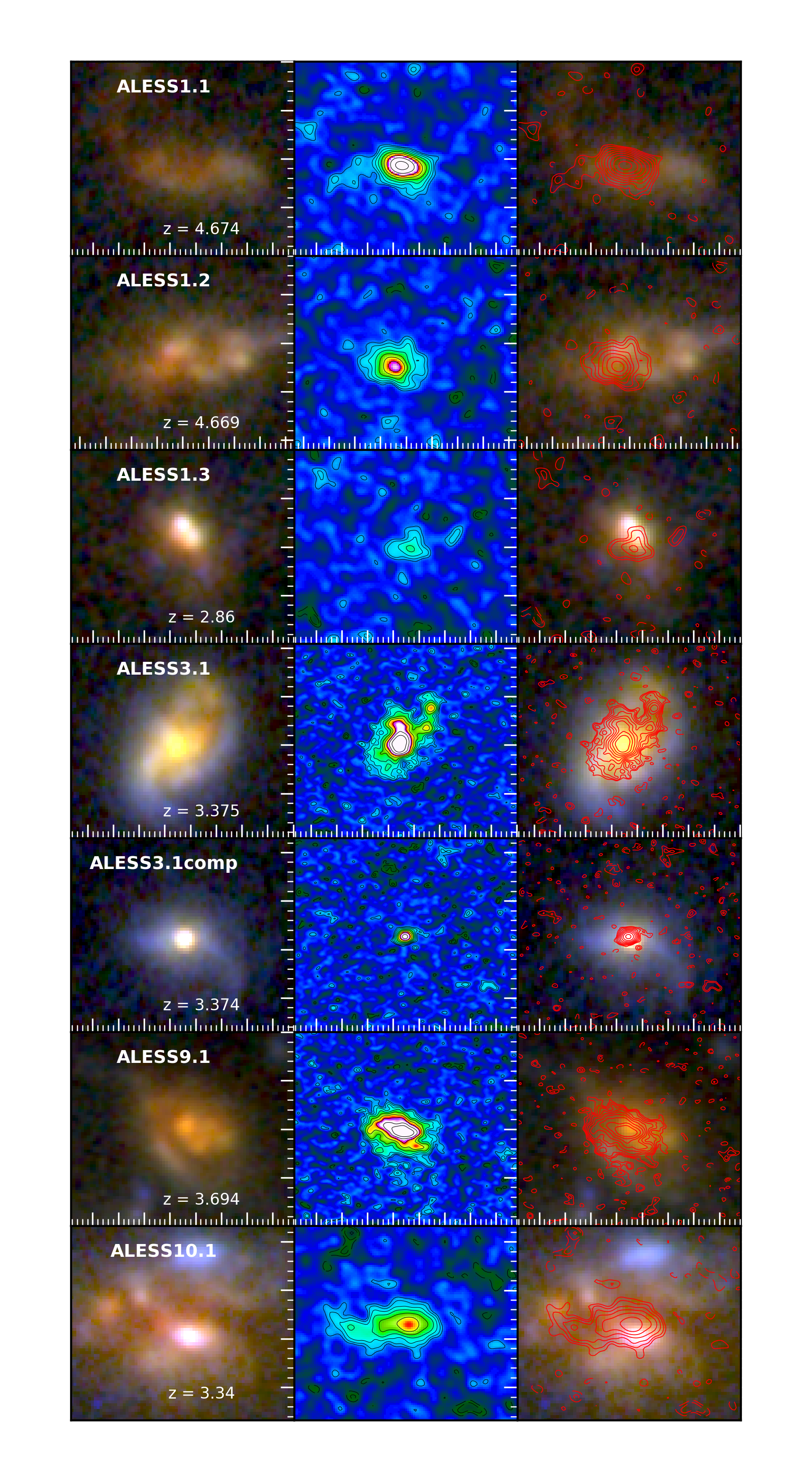}
\hfill
\includegraphics[trim={1cm 1cm 1cm 1cm},clip, width=0.49\textwidth]{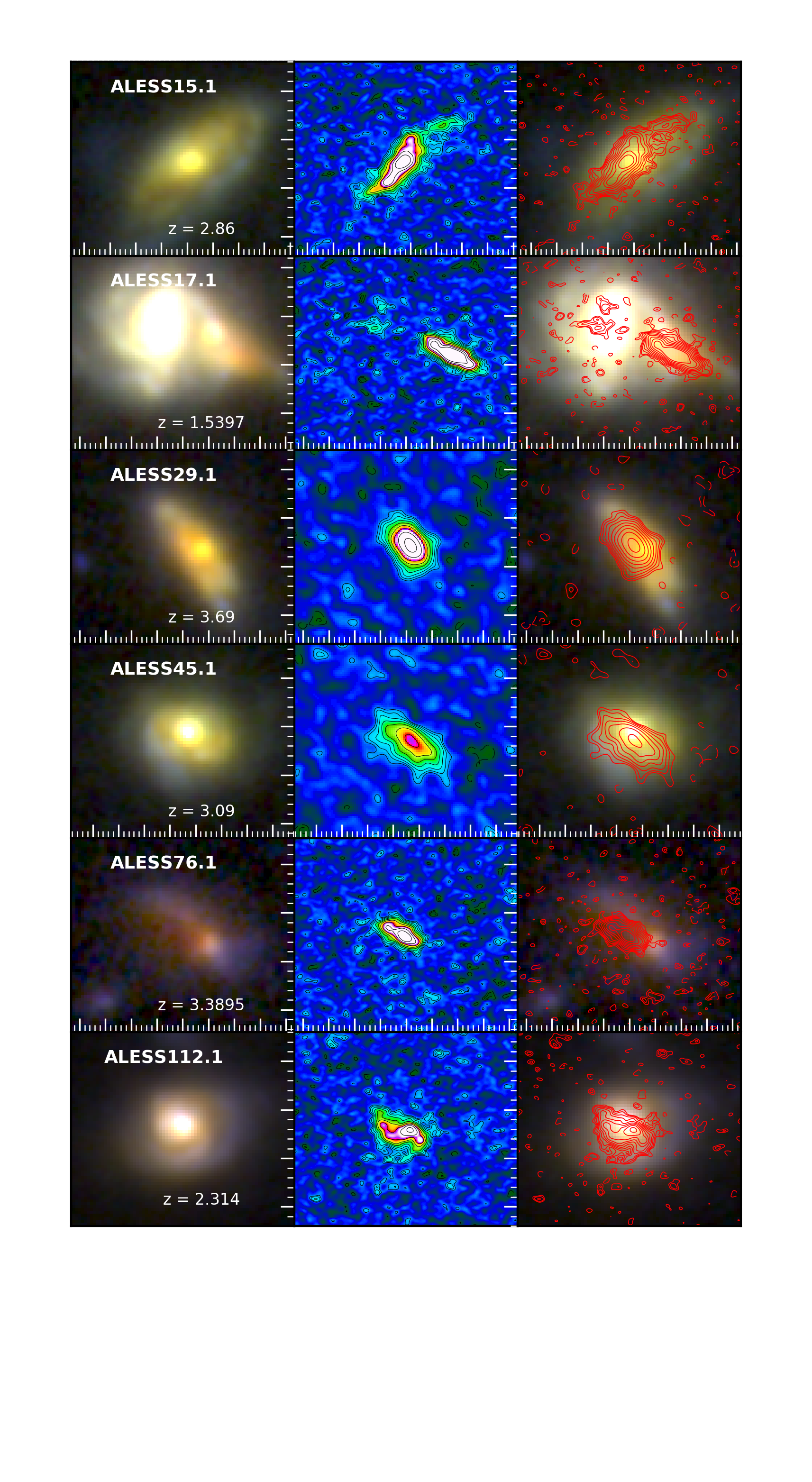}
\caption{JWST NIRCam versus ALMA 870$\mu$m imaging for the 
13 SMGs. For each source, the columns show the NIRCam RGB image (constructed from F444W, F356W, and F200W; left), the ALMA 870$\mu$m image 
(middle), and the ALMA contours in red overlaid on the RGB image (right). The ALMA images are scaled with a power-law stretch between -10$\sigma$ and 20$\sigma$ (exponent $=$ 2.1), with ALMA contours starting at $\pm$2$\sigma$ and increasing in powers of $\sqrt{2}$. 
Panels are 2$''$$\times$2$''$, or $\sim$15\,kpc at $z\sim3$. The NIRCam images reveal the rest-frame near-infrared morphologies of the galaxies on the same (sub-)kpc scales as the 870$\mu$m dust continuum, with the 870$\mu$m emission tracing the dust-reddened centers as well as some more extended features. 
}
\label{fig:JWSTvsALMA}
\end{figure*}

\section{Analysis \& Results}
\label{sec:analysis}

\subsection{Stellar emission compared to dust continuum}
\label{sec:JWSTvsALMA}

Fig.~\ref{fig:RGBfields} shows 8$''$$\times$8$''$ ($\sim$65 $\times$ 65\,kpc at $z\sim3$) three-color (`RGB') images of the 13 primary targets. The RGB images were made with the NIRCam F444W (red), F356W (green), and F200W (blue) filters. The stellar counterparts to all of the SMGs are visible in the NIRCam RGB images. Interestingly, the NIRCam imaging also reveals 
an apparent tidal connection between ALESS 1.1 and 1.2 (which have spectroscopic redshifts of $z = 4.674$ and $z=4.669$, respectively), providing the first direct evidence for an ongoing interaction. We note that while ALESS 1.3 also lies nearby, its photometric redshift ($z$ $=$ 2.86$^{+0.46}_{-1.53}$) 
indicates it may be at lower redshift.

Also apparent in Fig.~\ref{fig:RGBfields} is a 
galaxy $\sim$5$''$ south of ALESS 3.1 
which was not detected at 870$\mu$m in the original ALESS survey \citep[][]{Hodge_2013} but was strongly detected in the subsequent much deeper (and high-resolution) follow-up ALMA imaging utilized in this work, with an integrated flux density of $S_{\rm 870}$ $=$ 1.07 $\pm$ 0.06 mJy. (This is fainter than the 3.5$\sigma$ limit of the original ALESS map which had $\sigma_{\rm 870}$ $=$ 0.41\,mJy beam$^{-1}$; \citealt{Hodge_2013}). 
ALESS 3.1-comp is only $\sim$40 km s$^{-1}$ offset from ALESS 3.1 (based on a recent CO(5--4) detection; Westoby et al.\,in prep) and has a projected nuclear separation with ALESS 3.1 of $\sim$40 kpc. Given this, its strong submillimeter emission, and the tidal features apparent in the NIRCam images (Fig.~\ref{fig:NRCfilters}), it is possible that this galaxy is undergoing an early-stage interaction/merger with ALESS 3.1. We therefore refer to it as `3.1-companion' (hereafter `3.1-comp') and include it in the primary sample in Table~\ref{tab:sample} and for the analysis in the remainder of the paper.

To investigate how the morphology of the sources changes with wavelength, Fig.~\ref{fig:NRCfilters} shows 3$''$$\times$3$''$ cutouts ($\sim$25 $\times$ 25\,kpc at $z\sim3$) of the 
13 sources in each of the NIRCam filters along with the corresponding RGB images.  All of the sources are detected in at least the F356W and F444W NIRCam filters
(i.e., longward of 3.5$\mu$m observed-frame), including the red stellar counterpart to 
ALESS 17.1, which lies $\sim$0.8$''$ from an optically bright spiral galaxy companion at the same redshift\footnote{SINFONI spectroscopic imaging previously indicated that the two galaxies lie at the same redshift \citep{Chen_2020}; this is further confirmed by recent high-resolution ALMA CO(5--4) imaging (Westoby et al.\,in prep.)}.  
The majority of the sources  
show much weaker and/or centrally suppressed emission in NIRCam's F200W filter compared to the F356W/F444W filters, highlighting the red galaxy centers. 
Some apparent tidal features are also visible; we indicate a few prominent examples in Fig.~\ref{fig:NRCfilters} in the filter where they are most evident.

\begin{figure}[t!]
\centering
\includegraphics[trim={0cm 0.3cm 1cm 2cm},clip, width=0.49\textwidth]{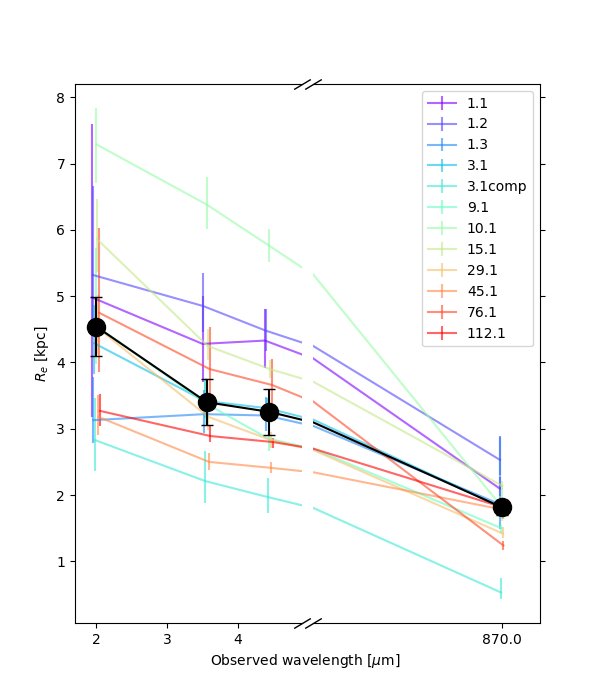}
\caption{Effective radius versus wavelength for the  
sources 
(excluding 17.1 due to confusion with  its optically bright companion). Individual sources are shown in different colors. The sample median is shown with the large black circles, where the error bars were calculated using bootstrapping. Including both the NIRCam and 870$\mu$m observations, we find that the median effective radius systematically decreases with increasing wavelength.
}
\label{fig:Re_vs_wavelength}
\end{figure}

Fig.~\ref{fig:JWSTvsALMA} shows a comparison of the JWST NIRCam RGB imaging (zoomed to 2$''$$\times$2$''$) and the high-resolution 870$\mu$m imaging for the  
13 sources. 
The morphologies of the (typically) rest-frame near-infrared stellar  and dust continuum images (traced by F444W and 870$\mu$m, respectively) show some striking similarities (as well as some notable differences), which we explore further below.
Here, we note that the peak of the emission from the rest-frame near-infrared counterpart (measured from the peak pixel in the F444W filter) is consistent with the peak of the 870$\mu$m emission to within the F444W PSF FWHM of $<$0.14$''$($\sim$1 kpc) for the majority (9/13) of the galaxies: ALESS 3.1, 3.1-comp, 9.1, 10.1, 15.1, 17.1, 29.1, 45.1, and 112.1. 
For the remaining four sources (ALESS 1.1, 1.2, 1.3, and 76.1), the peak of the 870$\mu$m emission is $\geq$0.14$''$ from the peak of the apparent counterpart in the F444W imaging, with the largest offset measuring 3 kpc (for ALESS 1.3, where the 870$\mu$m emission may be more consistent with the fainter of what appear to be two marginally blended components in the NIRCam images). 
For the sources with significant offsets, this may indicate either that the ALMA 870$\mu$m and NIRCam imaging are tracing physically distinct components in these sources, or that the detected stellar emission traced by the NIRCam F444W filter is still experiencing significant dust attenuation.
We explore these possibilities further below.

\subsection{Visual Classification}
\label{sec:classification}

We next visually classify the galaxies into three broad classes: 
\begin{description}
\item[M)] `Merger': Sources with evidence for a merger or interaction;
\item[I)] `Indeterminate';
\item[D)] `Disk': Sources with no clear evidence for a merger or interaction, and which thus appear as undisturbed disks.
\end{description}

To determine the visual classification for each galaxy, five coauthors independently classified the morphology of each of the 13 sources, taking into account the rest-frame near-infrared and 870$\mu$m morphologies (and further informed by the presence of a companion at the same redshift, if applicable). For each galaxy, we adopt the modal classification (preferred by at least 3/5); if no majority classification resulted from this method, then the galaxy was automatically assigned class I (indeterminate). We then calculated the fraction of sources in each of the three classes. Given the subjective nature of such classifications, we take the difference between the fraction in a given class from the above method and that from the most discrepant individual classification by any one classifier and adopt it as a rough estimate of the uncertainty. 

With the above method, we find the following results: M: 54$\pm$23\% (show evidence of mergers); I: 23$\pm$31\% (indeterminate); D: 23$\pm$8\% (appear as undisturbed disks). These results indicate that the majority of the galaxies in the current sample show signs of ongoing major/minor mergers or interactions, while only a minority show no clear evidence from the current datasets. 
Larger samples of galaxies spanning a wider range of 870$\mu$m flux density \citep[e.g.,][]{Gillman_2024} will be required to determine the importance of mergers/interactions for the SMG population as a whole.

\subsection{Flux density profiles}
\label{sec:lightcurves}

To determine the effective radii of the galaxies in each filter, we create curves of growth. 
We first  
run {\sc SExtractor} on the NIRCam images with a detection threshold of 3.0 times the local noise, a minimum number of pixels above the threshold of 12 at 0.03$''$ sampling per pixel, 
and 
DEBLEND$\_$MINCONT set to 0.05. The latter parameter prevents deblending ALESS1.2 and ALESS10.1 into two sources each. 
We determine the aperture shape for each source using the results from the F444W filter (chosen to minimize extinction), with the centroid taken as the peak pixel in 870$\mu$m, which will be more robust against dust-obscuration effects than the NIRCam filters.
We then use {\sc PhotUtils} to measure the flux density in elliptical apertures as a function of distance along the major axis, with the total integrated flux density in each filter taken 
using the F444W maximum aperture (again to minimize extinction effects, and after ensuring the emission from all filters is covered). 
Finally, we use the same apertures to measure the flux density in the 870$\mu$m image for each source. 

Fig.~\ref{fig:Re_vs_wavelength} shows the effective radii (and median values) derived from this method versus observed wavelength, and including both the NIRCam and 870$\mu$m measurements. Here we 
exclude ALESS 17.1 due to confusion with its optically bright companion in the JWST images. The values for individual sources can be found in Table~\ref{tab:morphology} 
and the flux density profiles themselves can be found in the Appendix. 

We find that the effective radius measured for the galaxies systematically decreases with increasing wavelength for the majority of the sources, with median values (and bootstrapped errors) of 0.61$''$$\pm$0.08$''$, 0.45$''$$\pm$0.06$''$, and 0.41$''$$\pm$0.05$''$ for the F200W, F356W, and F444W filters, respectively. These values correspond to 4.5$\pm$0.5\,kpc, 3.4$\pm$0.4\,kpc, and 3.0$\pm$0.3\,kpc using the individual source redshifts, with 16th--84th percentile ranges of 3.0--5.3\,kpc (F200W), 2.4--4.4\,kpc (F356W), and 2.4--4.2\,kpc (F444W). The effective radius then decreases even further for the majority of the sources at 870$\mu$m, with a median value of 0.23$''$$\pm$0.01$''$ (1.8$\pm$0.1\,kpc, with a 16th--84th percentile range of 1.4--2.1\,kpc). 
Overall, the median effective radius of the galaxies at F200W [F356W] is 48$\pm$30\% [10$\pm$20\%] larger 
than that measured at F444W, reflecting the relative brightness of the red galaxy centers at the reddest wavelengths. We explore whether this trend is due to dust obscuration in the following sections. Meanwhile, the median effective radius at F444W is 78$\pm$21\% larger than that measured with ALMA at 870$\mu$m. 
We note that if we had instead used the peak F444W pixel positions as the aperture centers, our conclusions would not change. We thus find much more compact dust continuum than rest-frame $\sim$1$\mu$m light, consistent with earlier indications from select sources \citep[e.g.,][]{Chen_2022, Gillman_2024}.

\begin{figure*}
\centering
\includegraphics[trim={1cm 1cm 1cm 1cm},clip, width=0.49\textwidth]{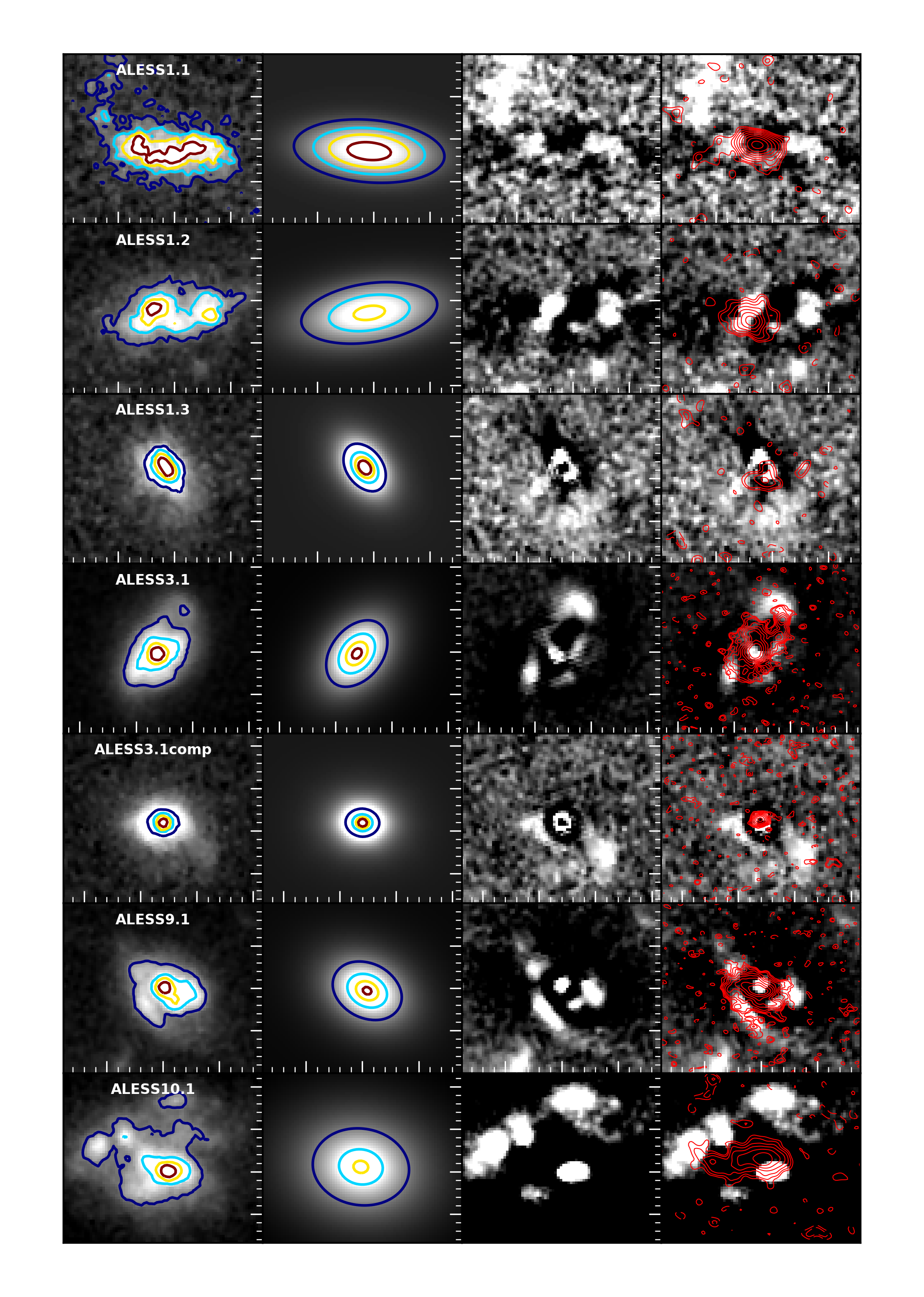}
\hfill
\includegraphics[trim={1cm 1cm 1cm 1cm},clip, width=0.49\textwidth]{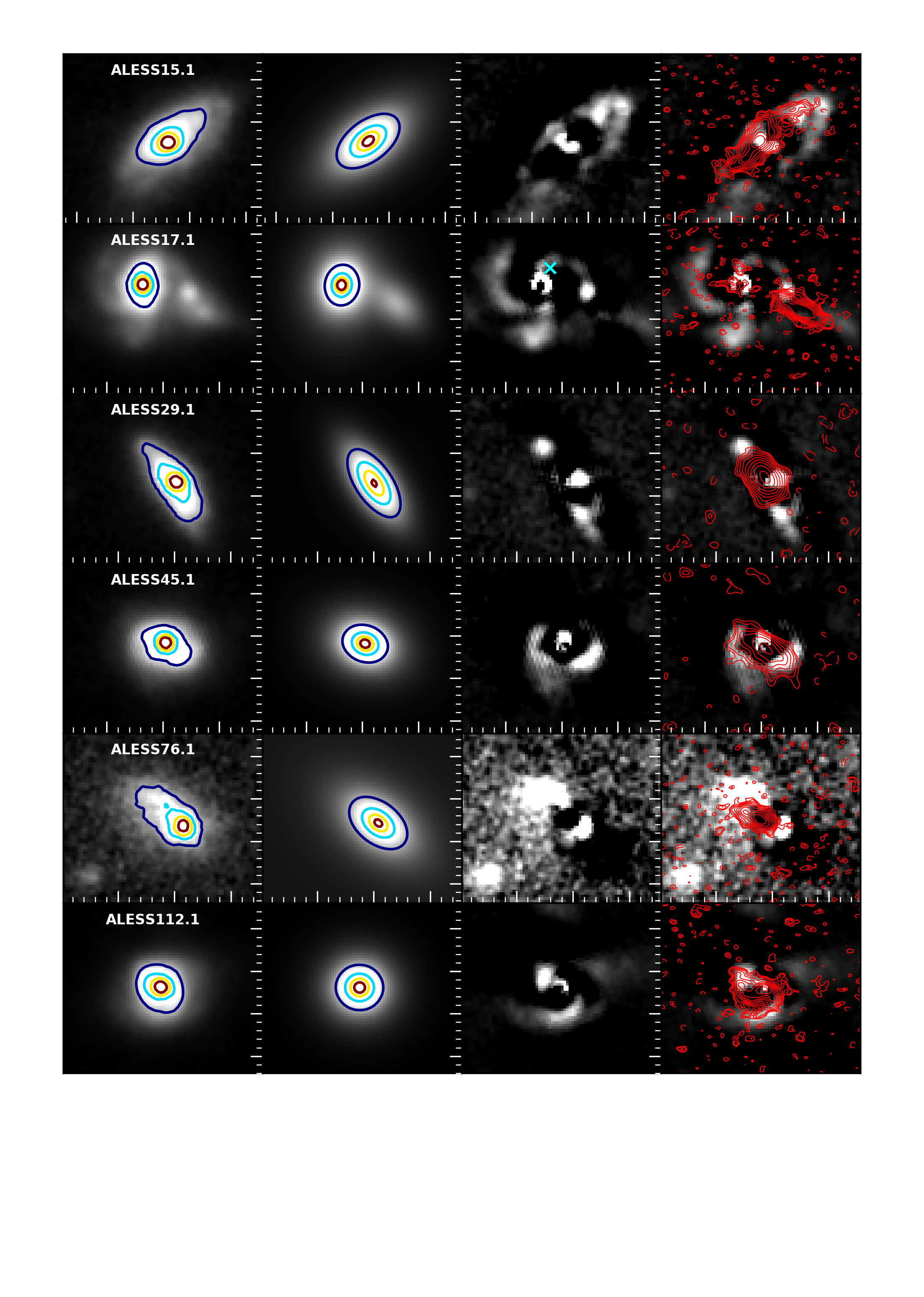}
\caption{{\sc Galfit} modeling of the NIRCam F444W images for the 
13 targeted SMGs. For each source, the columns show the NIRCam F444W image (1st column), the best-fit S\'ersic profile model (2nd column), 
 the residuals (3rd column), as well as a comparison of those residuals to the ALMA 870$\mu$m contours overlaid in red (4th column). 
The first and second column are scaled with a linear stretch between 0.25\% and 99\%, with contours showing 20\%, 40\%, 60\%, and 80\% of the peak intensity of the data (i.e., column 1). The third and fourth columns are shown with a linear stretch and the peak intensity scaled down by a factor of 5 to highlight the discrepancies. We note there is no qualitative difference to the residuals when the fits with freely varying $n$ values are shown instead. 
870$\mu$m contours start at $\pm$2$\sigma$ and increase in powers of $\sqrt{2}$, where $\sigma$$=$19$\mu$Jy. Panels are 2$''$$\times$2$''$, or $\sim$15\,kpc at $z\sim3$. See Section~\ref{sec:galfit} for details.  The cyan cross overlaid on the third column for ALESS 17.1 shows the position of the known X-ray AGN 
\citep[corrected for the median offset between the X-ray catalog and Gaia DR1;][]{Luo_2017}, indicating it is associated with the optically bright companion rather than the SMG.  
}
\label{fig:JWSTgalfit}
\end{figure*}

\begin{figure*}
\centering
\includegraphics[trim={0cm 0cm 0cm 0cm},clip, width=0.7\textwidth]{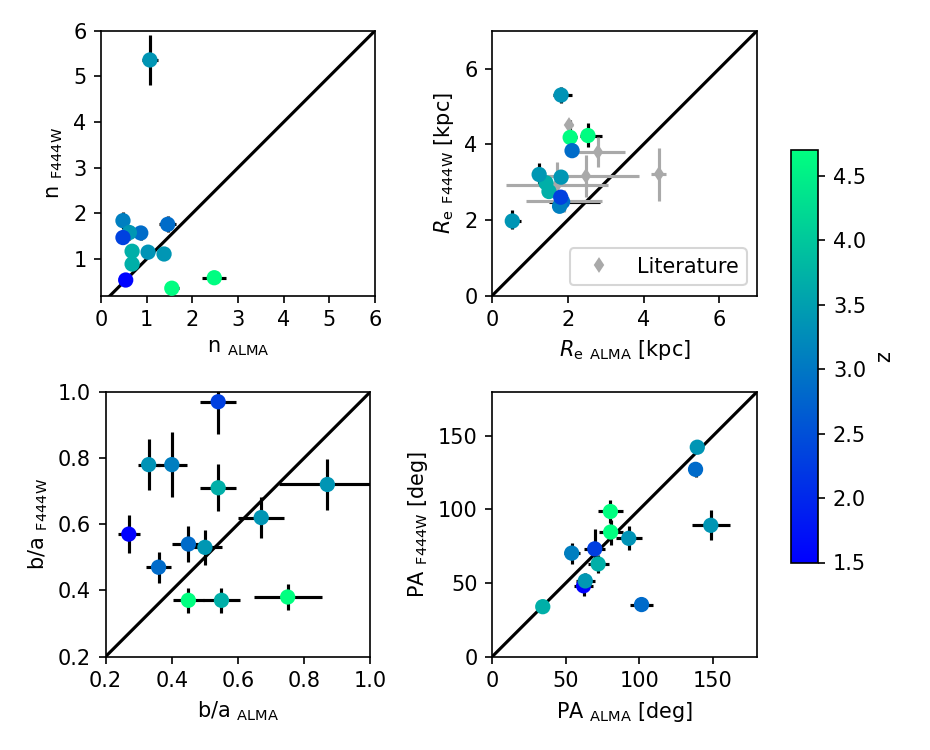}
\caption{Comparison of the morphological parameters derived from the F444W images with those from the 870$\mu$m images. The panels showing the S\'ersic index $n$ (\textit{upper left}), the axis ratio $b/a$ (\textit{bottom left}), and the major axis position angle \textit{{\rm (PA}, bottom right}) were derived using {\sc Galfit} with single S\'ersic profile fits (Section~\ref{sec:galfit}).  
For the axis ratio and position angle, the S\'ersic index is fixed to $n=1$ to faciliate direct comparison. The effective radii $R_{\rm e}$ (\textit{upper right}) were derived non-parametrically using curves of growth (Section~\ref{sec:lightcurves}). All data points are color-coded by redshift. The literature SMGs in the top right panel show other sources with measurements at both wavelengths (compiled from \citealt{Chen_2022, Cheng_2023, Kamieneski_2023, Smail_2023}). Comparing the F444W- and 870$\mu$m-derived parameters, we find a strong positive correlation along the one-to-one line between the position angles, a tentative positive correlation between the effective radii, and no correlation between either the S\'ersic indices or axis ratios.}
\label{fig:galfitparams}
\end{figure*}

\subsection{{\sc Galfit} modeling}
\label{sec:galfit}

To compare the morphology of the NIRCam-detected stellar emission with that of the 870$\mu$m dust continuum in more detail, we fit both the F444W images (chosen to minimize dust extinction) and the 870$\mu$m images 
with single two-dimensional S\'ersic profiles 
using {\sc Galfit} \citep{Peng2002, Peng2010}. We consider only single-component profiles for a more straightforward comparison  
between parameters. For the 870$\mu$m image fitting, we use the $\sim$0.16$''$ 870$\mu$m ALMA maps available for all of our targeted galaxies and verify that we recover the results of \citet{Hodge_2016}, noting that \citet{Hodge_2019} found S\'ersic indices that agreed within the errors for sources also observed at the higher resolution (0.08$''$) with ALMA. 

For the fitting of the F444W images, we obtain the most realistic PSFs by adopting the following strategy: We first use {\sc WebbPSF} v1.2.1 \citep{Perrin14} to generate the F444W PSF  
oversampled by a factor of four. We then match the simulated PSF to our science data for each target using the \texttt{webbpsf.setup$\_$sim$\_$to$\_$match$\_$data} utility. Finally, to further improve these models, we follow \citet{Chen_2022} and use {\sc Galfit} to convolve the resulting PSFs with Gaussian profiles and fit them to nearby (unsaturated) stars. This results in reduced $\chi^2$ values of $\sim$1, and we therefore adopt the best-fit (normalized) models as the final PSFs. 
We then run {\sc Galfit}, masking the other sources in the field using the segmentation maps from {\sc SExtractor} (and with the parameters presented in Section~\ref{sec:lightcurves}). 
For initial parameter guesses, we use the parameters derived from the 870$\mu$m image fitting, but we checked that our results were not sensitive to the exact input parameters.
For ALESS 17.1, we simultaneously fit the sMG and its optically bright companion.   

Fig.~\ref{fig:JWSTgalfit} shows the results of running {\sc Galfit} on the F444W images (where the values of the resultant S\'ersic indices, axis ratios, and position angles can be found in Table~\ref{tab:morphology}). 
From the residual panels, it is evident that many sources contain structures beyond a single S\'ersic profile. For example, 
roughly half of the sources (ALESS 3.1, 3.1-comp, 9.1, 15.1, 29.1, and 45.1) 
show evidence  
for excess emission beyond the models that is concentrated near the galaxy centers. 
A comparison of the {\sc Galfit} residuals with the high-resolution 870$\mu$m imaging shows that 
this excess emission is 
aligned with the peak of the 870$\mu$m emission in the majority of the cases.
These central residuals are thus potentially evidence for bulge-like components and/or bright central AGN, as also reported in JWST imaging studies of submillimeter-fainter sources \citep[e.g.,][]{Chen_2022}.

Beyond these central stellar concentrations, 
there are additional residual structures evident in Fig.~\ref{fig:JWSTgalfit} that are not captured in the {\sc Galfit} models.  
In particular, we see 
excess emission in the outskirts of 
several of the sources,
further highlighting some of the features already apparent in Fig.~\ref{fig:NRCfilters} resembling tidal tails/plumes (e.g., see ALESS 3.1 and 3.1-comp). Interestingly, some of these potential tidal features are also partially detected in the 870$\mu$m continuum (see Fig.~\ref{fig:JWSTgalfit}) for the sources with the highest-resolution 870$\mu$m imaging (e.g., ALESS 3.1, 9.1, and 112.1, though we note that the classification of these features can be somewhat subjective; Section~\ref{sec:classification}).

\begin{deluxetable*}{@{\extracolsep{4pt}}lccccccccccc@{}} 
	\tablecolumns{10}
	\tabletypesize{\footnotesize}
	\tablecaption{Morphology with JWST and ALMA}
	\tablehead{
         &
        \multicolumn{4}{c}{NIRCam F444W} &
        \multicolumn{4}{c}{ALMA 870$\mu$m} &
         & \\
        \cline{2-5}\cline{6-9}
        \colhead{Source ID} &
        $n$\tablenotemark{a} &
        $b/a$\tablenotemark{a} &
        PA\tablenotemark{a} &       
        $R_e$\tablenotemark{b} &
        $n$\tablenotemark{a} &
        $b/a$\tablenotemark{a} &
        PA\tablenotemark{a} &       
        $R_e$\tablenotemark{b} &
         Classification\tablenotemark{c} \\
        &
        & 
        &
        (deg) &
        (kpc) &
        &
        &
        (deg) &
        (kpc) 
        &
	} 
	\startdata
ALESS 1.1 & 0.36$\pm$0.04 &  0.37$\pm$0.04 & 85$\pm$8 &  4.2$^{+0.4}_{-0.4}$ & 1.6$\pm$0.2 & 0.45$\pm$0.05 & 80$\pm$8 & 2.1$^{+0.2}_{-0.1}$ & M \\
ALESS 1.2 & 0.59$\pm$0.06 & 0.38$\pm$0.04 & 98$\pm$8 &  4.2$^{+0.3}_{-0.3}$ & 2.5$\pm$0.3 & 0.75$\pm$0.10 & 80$\pm$9 & 2.5$^{+0.4}_{-0.2}$ & M \\
ALESS 1.3 & 1.8$\pm$0.2 & 0.54$\pm$0.05 & 36$\pm$4 &  2.5$^{+0.3}_{-0.2}$ & 1.5$\pm$0.2 & 0.45$\pm$0.05 & 102$\pm$8 & 1.9$^{+1.0}_{-0.4}$ & I \\
ALESS 3.1 & 1.1$\pm$0.1 & 0.62$\pm$0.06 & 142$\pm$4 &  3.1$^{+0.2}_{-0.1}$ & 1.4$\pm$0.1 & 0.67$\pm$0.07 & 139$\pm$4 & 1.82$^{+0.08}_{-0.08}$ & M \\
ALESS 9.1 & 1.2$\pm$0.1 & 0.71$\pm$0.07 & 62$\pm$6 &  2.8$^{+0.2}_{-0.1}$ & 0.68$\pm$0.07 & 0.54$\pm$0.05 & 72$\pm$7 & 1.50$^{+0.05}_{-0.07}$ & I \\
ALESS 10.1 & 1.2$\pm$0.1 & 0.78$\pm$0.08 & 80$\pm$8 &  5.3$^{+0.2}_{-0.2}$ & 1.0$\pm$0.1 & 0.33$\pm$0.03 & 93$\pm$9 & 1.8$^{+0.3}_{-0.2}$ & M \\
ALESS 15.1 & 1.6$\pm$0.2 & 0.47$\pm$0.05 & 127$\pm$5 &  3.8$^{+0.1}_{-0.1}$ & 0.87$\pm$0.08 & 0.36$\pm$0.04 & 138$\pm$4 & 2.14$^{+0.07}_{-0.07}$ & D \\
ALESS 17.1\tablenotemark{d} & 0.54$\pm$0.05 & 0.57$\pm$0.06 & 52$\pm$8 &  -- & 0.54$\pm$0.05 & 0.27$\pm$0.03 & 62$\pm$6 & -- & M \\
ALESS 29.1 & 0.89$\pm$0.09 & 0.37$\pm$0.04 & 34$\pm$3 & 3.0$^{+0.1}_{-0.1}$ & 0.68$\pm$0.07 & 0.55$\pm$0.06 & 34$\pm$3 & 1.42$^{+0.10}_{-0.07}$ & D \\
ALESS 45.1 & 1.8$\pm$0.2 & 0.78$\pm$0.10 & 69$\pm$7 & 2.4$^{+0.1}_{-0.1}$ & 0.48$\pm$0.05 & 0.40$\pm$0.04 & 54$\pm$6 & 1.8$^{+0.2}_{-0.1}$ & D \\
ALESS 76.1 & 1.6$\pm$0.2 & 0.53$\pm$0.05 & 52$\pm$5 & 3.2$^{+0.3}_{-0.2}$ & 0.62$\pm$0.06 & 0.50$\pm$0.05 & 63$\pm$6 & 1.24$^{+0.07}_{-0.07}$ & M \\
ALESS 112.1 & 1.5$\pm$0.2 & 0.97$\pm$0.10 & 81$\pm$10 & 2.6$^{+0.1}_{-0.1}$ & 0.48$\pm$0.05 & 0.54$\pm$0.05 & 70$\pm$7 & 1.81$^{+0.12}_{-0.09}$ & I \\
\hline
ALESS 3.1-comp & 5.4$\pm$0.5 & 0.72$\pm$0.08 & 88$\pm$9 & 2.0$^{+0.3}_{-0.2}$ & 1.1$\pm$0.2 & 0.87$\pm$0.14 & 148$\pm$13 & 0.53$^{+0.22}_{-0.09}$ & M 
    \enddata
    \tablenotetext{a}{Derived from S\'ersic fitting using {\sc Galfit} (Section~\ref{sec:galfit}).}
    \tablenotetext{b}{Derived from flux density profiles with {\sc PhotUtils} (Section~\ref{sec:lightcurves}).}
    \tablenotetext{c}{Classifications are M) Sources with evidence for a merger/interaction; I) Indeterminate; and D) Sources with no clear evidence for a merger/interaction, and which thus appear as undisturbed disks. See Section~\ref{sec:classification} for further details. }
    \tablenotetext{d}{The parameters reported here are for the SMG, not the optically bright companion.}
\label{tab:morphology}
\end{deluxetable*}

Fig.~\ref{fig:galfitparams} shows a comparison of the morphological parameters derived for each galaxy from the F444W and 870$\mu$m images. (The optically bright companion to ALESS 17.1 is not included in this plot due to its low-S/N in the 870$\mu$m image; see Fig.~\ref{fig:JWSTvsALMA}). 
The S\'ersic index ($n$), ellipticity ($b/a$), and position angle (PA) were taken from the best-fit {\sc Galfit} models.  
As the uncertainties reported by {\sc Galfit} are unrealistically small, we generate more realistic error bars by 
adding 10\% fractional uncertainty in quadrature on all parameters. 
In addition, we fix the S\'ersic indices $n$ to 1.0 for both F444W and 870$\mu$m to ensure a more straightforward comparison when fitting the other relevant structural parameters ($b/a$ and PA). We take the difference between the parameters calculated with $n$ held fixed and allowed to vary and add this in quadrature as an additional systematic uncertainty for these parameters.

Fig.~\ref{fig:galfitparams} also shows a comparison of the best-fit effective radii ($R_{\rm e}$) measured in the 870$\mu$m and F444W images using the curve of growth analysis of Section~\ref{sec:lightcurves}. For the F444W images, the $R_{\rm e}$ values derived from {\sc Galfit} are consistent with the curve-of-growth values within the uncertainties. For the 870$\mu$m images, the two methods also produce consistent results for all sources except ALESS 1.1, where {\sc Galfit} prefers (at $>$3$\sigma$) a smaller radius (0.9$\pm$0.1\,kpc). Given that the (non-parametric) curve of growth analysis is less sensitive to morphological asymmetries, we consider these values the fiducial values. However, we note that if we instead used the {\sc Galfit}-derived results, it would only strengthen our finding that the dust continuum is more compact than the rest-frame $\sim$1 $\mu$m (Section~\ref{sec:lightcurves}).

From Fig.~\ref{fig:galfitparams}, we see that some of the morphological parameters describing the dust continuum and detected stellar distributions are more tightly correlated than others. To quantify this effect, we calculate the Pearson correlation coefficient, $\rho$, between the F444W- and 870$\mu$m-derived parameters, which measures the strength of the linear monotonic correlation between two sets of data\footnote{We note that the PA is a cyclical quantity. However, as 1) we have unwrapped the PA values, and 2) the two PAs are not dictated by random processes, a linear correlation coefficient is sufficient.}. We find values of $\rho$ (and corresponding probabilities) of $-$0.14 (0.65), $+$0.68 (0.01), $-$0.02 (0.94), and $+$0.72 (0.005) for $n$, $R_{\rm e}$, $b/a$, and PA, respectively\footnote{We note that we have also calculated the Spearman's correlation coefficients,  
finding values of $\rho$ (and corresponding probabilities) of $-$0.28 (0.36), 0.65 (0.02), $-$0.09 (0.76), and 0.69 (0.009) for $n$, $R_{\rm e}$, $b/a$, and PA, respectively. The use of the Spearman's correlation coefficients instead thus does not change our conclusions.}.

This exercise indicates that the position angle is the morphological parameter that shows the strongest correlation (of those tested) between the F444W and 870$\mu$m images. 
Moreover, as Fig.~\ref{fig:galfitparams} shows, the position angle is not just strongly correlated, but it is correlated along the one-to-one line. 
This result indicates that the global morphologies of the rest-frame infrared and dust continuum emission are generally well-aligned. 
This close agreement 
can be attributed to the fact that the deep 870$\mu$m imaging -- while more compact than the F444W emission for these galaxies -- nevertheless still traces rest-frame near-infrared structures beyond the central bulges(/bars). In particular, the global PA agreement seen here appears to be driven in at least some cases by the (apparent) tidal features detected in the outskirts of some sources in both the 870$\mu$m emission as well as in the detected stellar emission (e.g., Figure~\ref{fig:JWSTvsALMA}). 
Meanwhile, the two clear outliers from the PA comparison in Fig.~\ref{fig:galfitparams} are both understandable: ALESS 1.3 shows evidence for two marginally blended components in the NIRCam imaging (driving the PA), and the 870$\mu$m image of ALESS 3.1-comp is relatively shallow and compact compared to NIRCam. 

\begin{figure*}[t]
\centering
\includegraphics[trim={0cm 0cm 0cm 0cm},clip, width=1.0\textwidth]{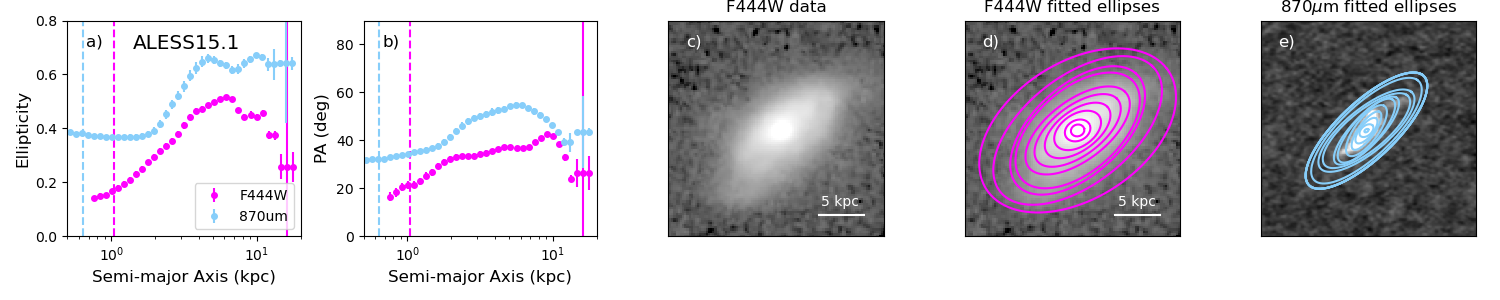}
\caption{An example of the ellipse fitting in the F444W filter to ALESS 15.1. Panels (a) and (b) show the ellipticity and PA (respectively) in F444W of the best-fit elliptical isophotes as a function of semi-major axis. The vertical dashed lines show the scale of the respective PSFs. (Note that the errors on the fits for the second-to-last ring extend beyond the ranges plotted.) The F444W data is shown in panel (c) with log scaling, while a linear selection of best-fit isophotes are overplotted in panel (d). Panel (e) shows the best-fit elliptical isophotes for the 870$\mu$m data for comparison. ALESS 15.1 is the only (non-clumpy) galaxy which meets all of the criteria for hosting a rest-frame near-infrared bar in the F444W imaging (see Section \ref{sec:bars}), but we consider this a tentative classification due to the combination of the long bar length implied, the 
marginally significant rotation of the PA in the outer disk, lack of evidence in the {\sc Galfit} residuals, and lack of PA correspondence between F444W and 870$\mu$m in the inner `bar' region. We therefore find no evidence for strong rest-frame near-infrared bars in the targeted galaxies. 
}
\label{fig:bar_search}
\end{figure*}

In addition to the strong correlation of the position angles, the effective radii show a tentative positive correlation. However, we also find -- perhaps surprisingly -- that the axis ratios show no correlation. This may be due to a real physical difference in the gas and stellar distributions, dust obscuration, or the fact that the axis ratios are tracing structure at smaller radii in the 870$\mu$m emission. 

We also find no correlation between S\'ersic indices in the F444W and 870$\mu$m emission. The Pearson's correlation coefficient suggests a mild anti-correlation, which may be caused by the influence of central dust obscuration on the F444W profiles, but it is not statistically significant. 
Finally, we note that the most extreme outlier from this plot, ALESS 3.1-comp, shows a very high S\'ersic index in the F444W image ($n=5.4\pm0.5$), which may hint at the presence of a bright compact nucleus (either a star cluster or a type 1 AGN) in the center.

\subsection{Isophotal fitting: Searching for stellar bars}
\label{sec:bars}

Given the previous evidence for bar-like features in the dust continuum in SMG samples \citep[e.g.,][]{Gullberg_2019}, and in particular, the elongated central features visible in the high-resolution images of some of the current targets \citep{Hodge_2019}, one might expect to find evidence for stellar bars. We thus search for potential stellar bars in the sources by fitting elliptical isophotes to the F444W images 
using the {\sc astropy} package {\sc PhotUtils}. Using this method, the surface brightness, ellipticity ($e = 1 - b/a$), and position angle of the fitted ellipses are allowed to vary as a function of distance along the semi-major axis, which increases by a factor of 1.1 for each subsequent ellipse fit. The fitting continues until either the relative error in the local radial intensity gradient is $>$0.5 for two consecutive ellipse fits or the outermost ellipse extends to the low signal-to-noise region (S/N $\leq$ 3). For comparison, we fit the highest-resolution 870$\mu$m data available for each galaxy. 

Following \citet{Guo_2023}, we carry out the ellipse fitting in a two-step process:  (1) In the first iteration, we keep the center of each fitted ellipse as a free parameter. (2) In the second iteration, we fix the center of all ellipses to the mean center found from fitting the inner six ellipses (i.e., central 0.3$''$) in step 1.

To ensure that any bars identified are reliable, we exclude highly inclined sources (with $i > 60\deg$ as inferred from the projected axis ratios of the outermost ellipse in F444W). This excludes only ALESS 1.1 and ALESS 1.2. We exclude also ALESS 17.1 due to confusion with its companion but include all ten remaining sources, including those that are clearly clumpy as a test of the robustness of the methodology.

We then test whether the resulting radial profiles meet the following criteria, which are characteristic of stellar bars \citep[e.g.,][]{Wozniak_1995, Jogee_2004, Marinova_2007}: 
\begin{enumerate}
\item The ellipticity ($e$) rises smoothly to a global maximum $e_{\rm max}$ $>$ 0.4 (where the position of the maximum then defines the bar end) before decreasing again in the outer disk. 
\item The position angle remains fairly constant ($\Delta$PA $<$20 deg) along the bar.
\item In the transition region between the bar and outer disk, $e$ drops by $>$0.1.
\item The position angle changes by $>$10 deg beyond the bar end.
\end{enumerate}
We note that our choice of $e_{\rm max}$ $>$ 0.4 conservatively restricts the selection to strong bars, 
as most ($>$70\%)  bars in near-infrared images of local spiral galaxies have ellipticities $>$ 0.4 \citep[e.g.,][]{Marinova_2007, Menendez-Delmestre_2007} and 
\citet{Jogee_2004} find that it is difficult to unambiguously identify weaker bars at $z>1$ \citep[c.f.,][]{Sheth_2008}. 
See Fig. 1 in \citet{Jogee_2004} for an example of radial profiles for a $z \sim 0.5$ galaxy displaying these characteristic bar signatures.

We find that no meaningful fit is possible for ALESS 10.1, which appears very clumpy/complex. The ellipse fitting also fails on the first or second iterations for ALESS 76.1 and 112.1 due to a strong dependence on the definition of the galaxy center. Given that these sources appear irregular (see the residuals from the GALFIT fitting in Fig.~\ref{fig:JWSTgalfit}), they will not be discussed further. 

Out of the remaining seven galaxies, only two sources meet all of the above criteria: ALESS 9.1 and ALESS 15.1. ALESS 9.1 is clearly clumpy, so we disregard it in the remaining discussion, except as a cautionary example that the criteria above are necessary, but not sufficient, to identify bars, particularly in sources with irregular morphologies. Indeed, we also find that the apparent tidal features in the outskirts of some sources (e.g., ALESS 3.1) can 
cause a second peak in the ellipticity at large radii, which can complicate bar identification using the above criteria, but we have verified that we do not find any additional bar candidates if we define the `outer disk' using isophotes at smaller radii. 

ALESS 15.1 is thus the only remaining bar candidate. The ellipse fitting results for this source are shown in Fig.~\ref{fig:bar_search}. If there is a bar in this source, the ellipse fitting suggests it has a maximum ellipticity of $\sim$0.5
and a semi-major axis length of $\sim$6.3\,kpc. The latter value is within the range of bar lengths reported for $z<1$ galaxies \citep[e.g.,][]{Jogee_2004}
but on the large end of the range of values found for bars recently reported in $z>1$ galaxies with JWST \citep[$\sim$2.0--7.5\,kpc;][]{Guo_2023, Smail_2023, Huang_2023}. 
Given 1) that this galaxy is at higher redshift ($z_{\rm phot} = 2.67$) than all but one of those sources, where bar lengths should be shorter on average due to bar size evolution \citep[e.g.,][]{Debattista_2000, Martinez-Valpuesta_2006, Algorry_2017};  2) 
the marginal change in position angle beyond the `bar' end ($\sim$10 deg; i.e., barely above the threshold for a bar classification);
3) the lack of evidence for a bar in the {\sc Galfit} residuals; and 4) the lack of correspondence between the position angle of the F444W and 870$\mu$m isophotes in the `bar' region (Fig.~\ref{fig:bar_search}), where we would expect agreement between the two tracers in the case of a real bar; we consider 
this bar classification tentative at best.
We therefore find no convincing evidence for strong stellar bars in the current rest-frame near-infrared data. 

\begin{figure*}
\centering
\includegraphics[trim={0cm 0cm 0cm 0cm},clip, width=1.0\textwidth]{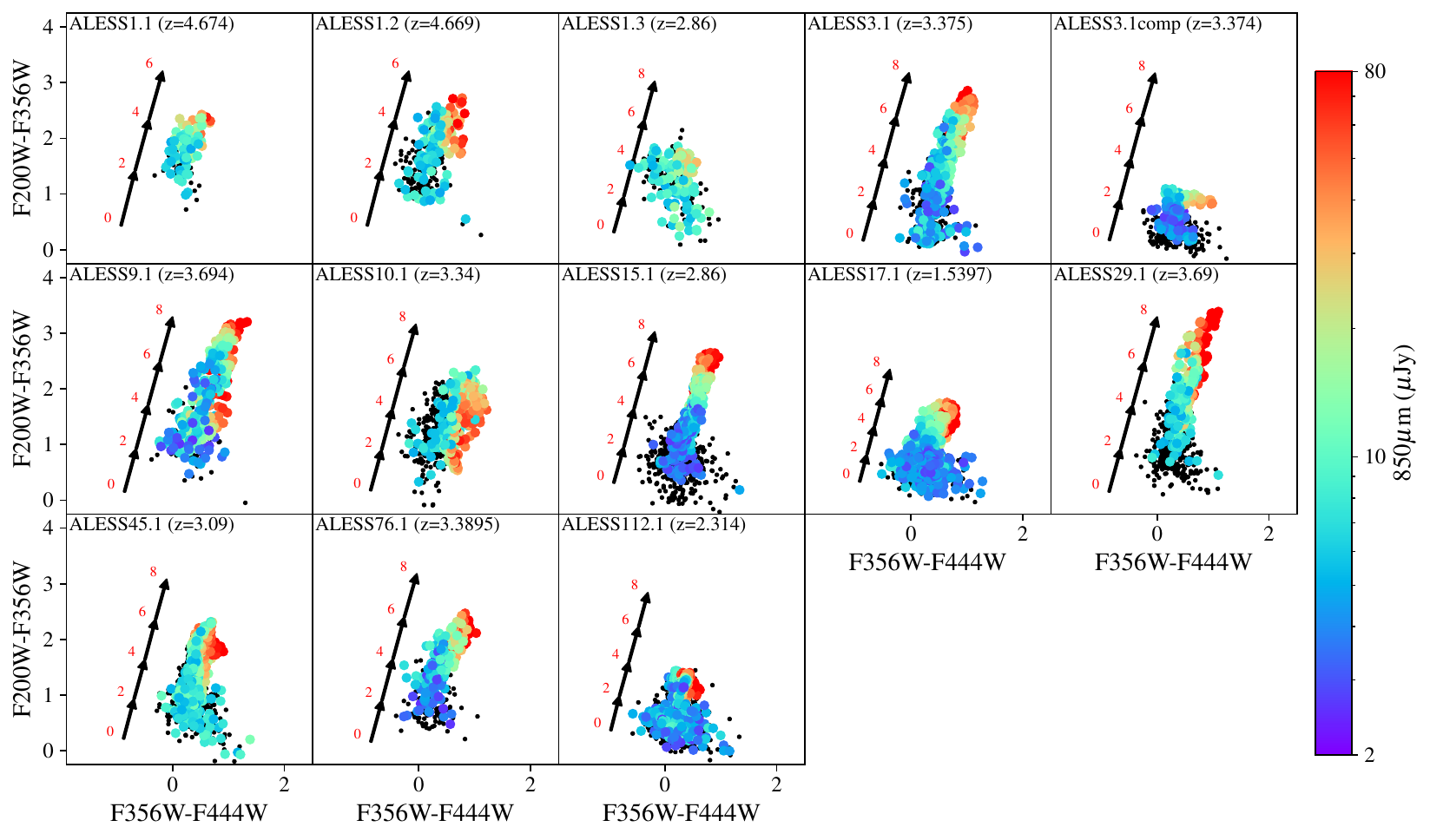}
\caption{NIRCam color-color diagrams showing F200W--F356W versus F356W--F444W colors for individual 60$\times$60\,mas regions within each galaxy with $>$5$\sigma$ detections in all three NIRCam filters. The data points are color-coded by their corresponding 870$\mu$m flux surface density if the aperture has at least a 1$\sigma$ detection in the ALMA 870$\mu$m map; otherwise the data points are colored black.
Vectors indicating the predicted impact on this color space of varying $A_{\rm V}$ are shown for comparison and were obtained with {\sc Magphys} \citep[][see text for details]{daCunha_2008, daCunha_2015}.
 The vectors are plotted with a shift 1\,dex to the left for clarity. The majority of the sources show a correlation between redder NIRCam colors and 870$\mu$m surface brightness. 
}
\label{fig:color-color}
\end{figure*}

\begin{figure*}
\centering
\includegraphics[trim={0cm 0cm 0cm 0cm},clip, width=0.8\textwidth]{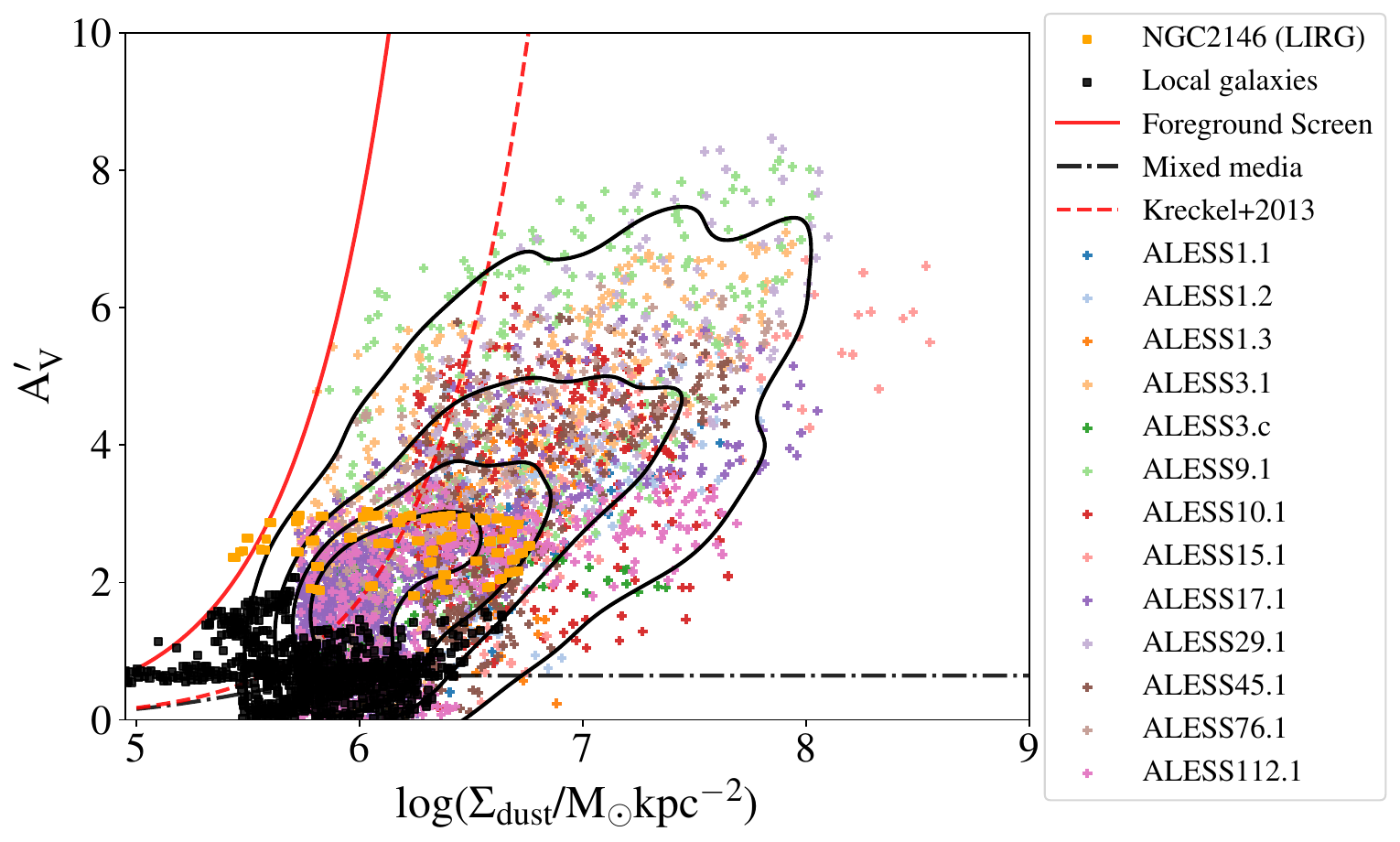}
\caption{V-band extinction (inferred from the reddening of the stellar continuum) as a function of dust mass surface density for the 13 SMGs and a comparison sample of local galaxies from \citet{Kreckel_2013}. 
The black contours show the density of SMG data points, which are shown in different color circles for each galaxy. The local galaxies are shown as black squares, except for NGC2146 (the only LIRG in the sample), which is shown as the orange squares. Overplotted are three lines: a foreground dust screen model (solid red line); a mixed media model (dot-dashed black line); and the empirical relation found by \citet{Kreckel_2013} for nearby galaxies (dashed red line; see their Equation 8) with the assumption of similar $A_{\rm V}$ estimates for SMGs from stellar reddening and the Balmer decrement \citep[][Taylor et al.\,in prep.]{Birkin_2022PhDT}. The distribution and scatter seen for the SMGs are similar to that seen for local galaxies while extending the trend to higher dust mass surface densities. See Section~\ref{sec:color-color} for further details. 
}
\label{fig:MdustvsAv}
\end{figure*}

\subsection{NIRCam color-color diagrams versus ALMA 870$\mu$m}
\label{sec:color-color}

Lastly, we quantitatively study the relationship between the 870$\mu$m dust continuum and detected stellar emission on $\lesssim$1\,kpc scales to relate the influence of the former on the latter and the effect that dust obscuration has on our understanding of high-redshift star-forming galaxies. 
The methodology we utilize will be discussed further in Li et al. (in prep.). In short, we first re-sample the 870$\mu$m images to match the coarser NIRCam pixel scale. We use a 1000$\times$1000 pixel grid with 30 mas pixels, fixing the coordinates of the target (Table~\ref{tab:sample}) at the centre of the central pixel. 
We then PSF-match the pixel-aligned NIRCam F200W, F356W and ALMA 870$\mu$m images to the F444W filter images using PSF models from {\sc WebbPSF} and 2D elliptical Gaussians for the NIRCam PSF and ALMA synthesized beam, respectively\footnote{We note that while the ALMA resolution is nominally slightly worse than that at F444W for 6/13 sources (0.16$''$ versus 0.14$''$), the PSFs have very different shapes. In particular, the F444W PSF has a more compact core but a much more extended diffraction pattern. We thus PSF-match to F444W for all sources, noting that 
the exact choice of kernel does not affect our conclusions.}. In particular, we generated a PSF-matching kernel for each image that requires convolution using the create$\_$matching$\_$kernel function in the {\sc PhotUtils} package. We then convolve these images with the PSF-matching kernel using the convolve function in the {\sc astropy} package after masking all other detected sources in the field of our targets.

To create color-color diagrams, we adopt an aperture grid with apertures of 2$\times$2 pixels (60 mas as these are Nyquist sampled). 
The fluxes within each aperture (i.e. resolution element) are summed and recorded in a table for all filters. The error on each measured flux density value was obtained by randomly placing 2$\times$2 pixel apertures in the unmasked background region in the images. 
For apertures with $\geq$3$\sigma$ detections in all three NIRCam filters, we compute the F200W--F356W and F356W--F444W colors and plot them in Fig.~\ref{fig:color-color}. The data points are color-coded by their corresponding 870$\mu$m flux density if the aperture has at least a 1$\sigma$ flux density in the ALMA 870$\mu$m map; otherwise the data points are colored black. In order to qualitatively assess the dust reddening properties in these sources, we also overplot vectors indicating the predicted impact on the colors of varying $A_{\rm V}$. We emphasize that these vectors are not fits to the data, but rather were obtained using {\sc Magphys} spectral libraries \citep{daCunha_2008, daCunha_2015} by finding the median color of stellar models within $\pm$0.1\,dex of a given $A_{\rm V}$ value, shown from $A_{\rm V}$ $=$ 0 to a maximum $A_{\rm V}$ that depends on redshift.

Fig.~\ref{fig:color-color} shows that for the majority of the sources, we see an extended distribution of points in the NIRCam color-color space, where the direction of the extension roughly matches that expected from the $A_{\rm V}$ vector. These distributions show that, as expected, the F200W--F356W color depends more strongly on $A_{\rm V}$ in the models than the F356W--F444W color (i.e., the vectors are closer to vertical than horizontal); this is due to the fact that dust attenuation affects shorter wavelengths more strongly than longer wavelengths. 
By comparing the observed distributions to these vectors, we find that the maximum $A_{\rm V}$ implied within each source on $\sim$1\,kpc scales (for regions detected in F200W) ranges from $A_{\rm V}$$<$4 in ALESS 3.1-comp (which is the only source below the detection threshold of the original ALESS catalog) to $A_{\rm V}$ $\sim$ 8 (in ALESS 9.1 and 29.1). For comparison, the median $A_{\rm V}$ of the ALESS sample from global spectral energy distribution (SED) fitting is only $A_{\rm V}$$\sim$2\,mag \citep{daCunha_2015}. 

Fig.~\ref{fig:color-color} also shows that for the majority of the sources, there is a correlation between redder NIRCam colors and 870$\mu$m surface brightness; i.e., the chance of a point being bright at 870$\mu$m is higher if it is red in both NIRCam colors. The clearest example of this correlation is seen in ALESS 15.1, but strong trends are also visible in ALESS 1.1, 1.2, 3.1, 9.1, 17.1, 29.1, 76.1, and 112.1 (and a weaker trend is visible in ALESS 1.3). This is a strong indicator that dust is the cause of the correlation. 

Finally, we see that for some sources, there is a large variation in F200W-F356W color for a fixed F356W-F444W color and 870$\mu$m surface brightness. ALESS 10.1 is the clearest example, but see also 1.2, 9.1, and 29.1.  While ALESS 10.1 is visually classified as a merger/interaction, the appearance of this effect in the larger sample does not correlate with morphological classification (nor does it correlate with, e.g., F444W ellipticity). It is possible that this is due to a `frosting' of low-$A_{\rm V}$ stars in front of the bulk of the higher-$A_{\rm V}$ emission. The full implications of these diagrams for the distribution of stellar age, (dust-corrected) stellar mass, metallicity, and other derived parameters will be explored in Li et al. (in prep.). 

To further explore the general relation between dust in absorption and dust in emission, in Fig.~\ref{fig:MdustvsAv} we plot $A_{\rm V}$ versus dust mass surface density for our sources. Here the $A_{\rm V}$ values were derived by taking each pixel in the color-color diagram of Fig.~\ref{fig:color-color} with a $>$1$\sigma$ 870$\mu$m flux density and projecting it onto the $A_{\rm V}$ vector at the source redshift (shown in Fig.~\ref{fig:color-color}). To emphasize that these values are not fits to the data but purely empirical, we refer to this axis as $A'_{\rm V}$ in the figure. For the dust mass surface density, we use the calibration between 870$\mu$m flux density and dust mass from \citet[][]{Dudzeviciute_2020}, where the dust masses were derived from {\sc Magphys}. We also compare to a sample of nearby galaxies from \citet{Kreckel_2013} with $A_{\rm V}$ values derived from the reddening of the stellar continuum. Finally, we overplot three relations: the relation for a simple foreground dust screen between the emitter and absorber (based on the dust model from \citealt{Draine_2007} and using the dust-to-gas ratio from \citealt{Draine_2014}); the relation for a mixed media model where stars and dust are uniformly mixed (from \citealt{Calzetti_2000} following \citealt{Kreckel_2013}); and the empirical relation established in \citet{Kreckel_2013} for their local galaxies \citep[with the assumption of similar $A_{\rm V}$ estimates for SMGs from stellar reddening and the Balmer decrement;][Taylor et al.\,in prep.]{Birkin_2022PhDT}. 
The \citet{Kreckel_2013} relation was determined 
on physical scales from 350\,pc to 2\,kpc, which is well-matched to the scales probed here ($\sim$1\,kpc),  
and it was derived by scaling the simple foreground dust screen model by a factor of $\sim$4 to account for geometric effects. 

Fig.~\ref{fig:MdustvsAv} shows that 
the relation between dust in absorption and dust in emission for the galaxies studied in this work broadly agrees with the trend reported for local galaxies, while also extending the trend to higher dust mass surface densities and V-band extinction. The scatter observed for the current targets also appears broadly consistent with that seen in the local sample, both between and within individual galaxies; in particular, we note that NGC2146 (the only LIRG in the nearby sample) shows a similar spread to some of our sources.  We discuss the additional implications of these findings in the following section.

\section{Discussion}
\label{sec:discussion}

\subsection{The influence of dust on rest-frame optical/near-infrared morphologies}
\label{sec:Discussion_dustinfluence}
The JWST NIRCam images presented here reveal the bulk of the underlying stellar emission in these high-resolution ALMA-mapped submillimeter-selected galaxies. 
For this, observing at or beyond 3.5$\mu$m (i.e., rest wavelength $\simeq$1$\mu$m at $z\sim3$) appears to be crucial. The morphologies can vary dramatically between F200W and F356W (i.e., between $\sim$0.5 and $\sim$1$\mu$m rest-frame), with the majority of the sources appearing either barely detected or with suppressed central emission at F200W compared to F356W (Fig.~\ref{fig:NRCfilters}). However, at F356W, all sources are well detected, including several galaxies which were previously undetected in deep HST images at $\leq$1.6$\mu$m (ALESS 1.1, 17.1, and 76.1; \citealt{Chen_2015}). 

The correspondence for the majority of sources between the morphology of the NIRCam F444W and the matched-resolution 870$\mu$m emission (particularly the peak positions and PAs) indicates that the spatial offsets previously reported between the 870$\mu$m continuum and HST 1.6$\mu$m emission \citep[e.g.,][]{Hodge_2012, Chen_2015, CalistroRivera_2018} were not due to true physical offsets, but rather that the morphologies inferred from the HST images were strongly affected by heavy structured dust obscuration, an effect also reproduced by radiative transfer modeling \citep{Cochrane_2019, Popping_2022}. 
For example, ALESS 29.1 was previously classified as a source with two separate stellar components, while four separate components were identified in ALESS 15.1 \citep{Chen_2015}. By uncovering the obscured centers of the underlying disks, the NIRCam images now clearly reveal that these `components' were simply due to differential dust obscuration in the disks. The higher $A_{\rm V}$ values in the galaxy centers (Fig.~\ref{fig:color-color}) are thus causing the observed trend of larger effective radius at decreasing NIRCam wavelength (Fig.\ref{fig:Re_vs_wavelength}).  We note that other studies have also reported spatial offsets between the rest-frame UV/optical stellar and ISM emission in a variety of high-redshift populations \citep[e.g.,][]{Cheng_2020, Cochrane_2021, Inami_2022, Killi_2024}, with various theories posited beyond dust geometry, including strong stellar feedback or large-scale gas inflow/outflows. While there may certainly be other mechanisms operating in different galaxy populations, we caution that---given the strong impact of dust demonstrated here up to $\simeq$1$\mu$m (rest-frame)---structured dust obscuration should be ruled out before invoking more exotic explanations.

For some of the highest-redshift sources, 
we see strong evidence that the galaxies are still affected by significant dust obscuration even in the longest-wavelength (F444W) NIRCam filter. 
In particular, examining the F444W versus 870$\mu$m continuum imaging for ALESS 1.1  ($z=4.674$; Fig.~\ref{fig:JWSTvsALMA}), as well as the comparison of the 870$\mu$m map with the F444W {\sc Galfit} residuals (Fig.~\ref{fig:JWSTgalfit}), the dust continuum appears to align with a suppression in the intensity of the detectable stellar distribution, suggesting significant dust obscuration. This suppression in the central region of the F444W image likely explains the very low F444W-derived S\'ersic index ($n=0.36\pm0.04$). 
We note that we find a similarly low S\'ersic index at F444W for our other $z>4.5$ source, ALESS 1.2. 
In these sources, F444W is probing shorter rest-wavelength emission (780\,nm rest-frame) that is more prone to dust extinction. 
For these particularly high-redshift/dusty sources, MIRI imaging will be important to more robustly characterize the stellar distributions.

\subsection{Implications for inclination estimates}
\label{sec:Discussion_inclinations}

While some of the most basic morphological parameters show agreement between the 870$\mu$m and F444W observations, others show no clear correlation (Fig.~\ref{fig:galfitparams}). The most notable parameter showing a lack of correlation is the axis ratio (i.e., ellipticity). This lack of correlation has significant implications, as the ellipticity is used to set the inclination in the most simple but commonly used dynamical mass estimates (often even without error margins), with the tacit assumption that these are perfectly flat, infinitely thin and perfectly circular disks. The lack of correlation between the ellipticites derived from different tracers shows that these inclinations are in fact very poorly determined, suggesting such dynamical mass estimates should be used with extreme caution. 

\subsection{Rest-frame near-infrared vs. dust continuum sizes}
\label{sec:Discussion_sizes}

The smaller extents measured for the ALMA 870$\mu$m emission compared to those inferred from the F444W emission (Fig.~\ref{fig:Re_vs_wavelength}) suggests that the 870$\mu$m emission is more weighted to the dusty centers of the sources. This is consistent with  
recent results inferred for JWST-observed SMGs compared with submillimeter continuum imaging of (largely) other samples \citep[e.g.,][]{Chen_2022}, but this study extends these results to direct comparisons of these tracers within the same galaxies. For sources 
without evidence for significant obscuration in the F444W filter, which could bias those size estimates to larger values, 
the smaller 870$\mu$m continuum sizes measured here would suggest that these galaxies are still undergoing rapid morphological transformation, with the central starbursts likely building bulges that will eventually come to dominate the stellar mass distributions (for which the F444W filter serves only as a weak proxy).
This result would strengthen the challenge these sources pose to recent hydrodynamical simulations, a number of which predict the opposite trend for galaxies in this stellar mass range \citep[i.e., submillimeter extents that are larger than the stellar mass extents; e.g.,][]{Cochrane_2019, Popping_2022}. Given the significant obscuration still present in even the longest-wavelength NIRCam images in at least some sources, this result needs to be tested via the creation of (dust-corrected) stellar mass maps (see \citealt{Smail_2023} and Li et al., in prep).

\subsection{Evidence for mergers and interactions}
\label{sec:discussion_mergers}

While the NIRCam observations have now detected the dust-reddened centers of the galaxies---suggesting less irregular morphologies than implied from the heavily obscured 1.6$\mu$m imaging \citep{Chen_2015}---this does not imply that the galaxies all appear as undisturbed disks in F444W. On the contrary, the NIRCam images reveal evidence for mergers/interactions in the majority of the sources. 

In particular, three of the SMGs have direct evidence for companions: ALESS 1.1 (with 1.2), ALESS 3.1, and ALESS 17.1. 
The most dramatic feature is the long tidal tail/stellar bridge that is now clearly visible between ALESS 1.1 and 1.2 at $z=4.67$ (Fig.~\ref{fig:RGBfields}). 
With a projected separation of $\sim$20 kpc and a velocity separation of $\sim$320 km s$^{-1}$, this suggests that ALESS 1.1 and 1.2 are undergoing an early-stage (violent) interaction. Interestingly, both sources are detected in CO(5-4) emission \citep{Birkin_2021}, where ALESS 1.1 has a double-peaked profile with a very wide FWHM velocity width of 1300 km s$^{-1}$. This 
could indicate that ALESS 1.1 is itself a late-stage merger or that the gas disk is dynamically disturbed. 
With ALESS 1.3 lying in the same field---and showing evidence for two very close components, where the southern component is redder (Fig.~\ref{fig:NRCfilters}) and may be the primary source of its submillimeter emission 
(Fig.~\ref{fig:JWSTvsALMA})---this means that the original submillimeter source `LESS 1'  (i.e., the brightest source in the original LABOCA `LESS' catalog, with S$_{\rm 850}$ $=$ 14.5 $\pm$ 1.2 mJy; \citealt{Weiss_2009}), was actually made up of at least two or more different and likely unrelated major merger events within its 19$''$ beam (FWHM). Meanwhile, while ALESS 3.1 and 3.1-comp are also likely undergoing an interaction, we note that ALESS 3.1 also has a broad (870 km s$^{-1}$) and asymmetric CO line profile \citep{Birkin_2021}, 
potentially indicating that ALESS 3.1 itself is undergoing a late-stage interaction/merger.
Thus the original submillimeter source `LESS 3' (i.e., the third brightest source in the LESS catalog) also appears to consist of at least one or more mergers/interactions, implying that the brightest LESS (i.e., single-dish LABOCA) sources are mergers/interactions\footnote{We note that LESS 2 was not included in the high-resolution ALMA/JWST follow-up due to the sample selection.} (and see \citealt{Mitsuhashi_2021} for a discussion of an overdensity associated with the brightest SCUBA-2 source in COSMOS). This is a direct confirmation of trends reported previously based on the multiplicity of sources comprising bright single dish-selected sources \citep[e.g.,][]{Stach_2018, Simpson_2020}.

The NIRCam images (and {\sc Galfit} residuals) also reveal evidence for interactions via features resembling stellar tidal tails/plumes and clumps in ALESS 10.1 and 76.1, as well as potential evidence in ALESS 9.1 and 112.1. 
Some of the apparent tidal features were previously detected in the dust continuum in the sources with the highest resolution 870$\mu$m emission maps (e.g., ALESS 3.1, 112.1; \citealt{Hodge_2019}) and interpreted as either tidally induced spiral arms or the star-forming knots in an interaction/merger. ALESS 3.1 was already discussed above and indeed shows evidence of a merger/interaction also in the NIRCam images. 
With the new NIRCam imaging showing possible evidence of a warped disk (Fig.~\ref{fig:JWSTgalfit}), ALESS 112.1 may also be a late-stage merger, though this case is less conclusive and is thus marked as Indeterminate. 
ALESS 9.1, which also has a potential tidal feature detected in both NIRCam and 870$\mu$m continuum (to its South), appears very clumpy in its rest-frame near-infrared emission, possibly due to an advanced interaction/merger (though again marked `Indeterminate' due to the lack of clear evidence). In fact, of the 13 sources examined, only ALESS 15.1, 29.1 and 45.1 
show no clear evidence in the current data for mergers/interactions, 
though we caution that these sources may still be late-stage mergers or interactions where the merger/interaction is less obvious due to projection effects or low-mass companions. 
Conversely, it should be noted that 
the departures from asymmetry used here to visually classify the galaxies as mergers/interactions in cases without a clear companion may also result from accretion from a cooling halo in gas-rich systems, which can be episodic and sporadic \citep[e.g.,][]{Bland-Hawthorn_2024}.

Keeping these caveats in mind, the current interpretations result in a merger fraction of 54$\pm$23\% and a fraction of undisturbed disks of only 23$\pm$8\%---consistent (within the errors) with the larger but typically submillimeter-fainter sample of \citet{Gillman_2024}, who report 20$\pm$5\% candidate late-stage mergers plus 40$\pm$10\% potential minor mergers. 
Confirming these interpretations will ultimately require high-resolution kinematic data. However, regardless of their nature, the fact that a number of these (apparent tidal) features are detected also in the submillimeter continuum indicates that---despite the centrally concentrated 870$\mu$m emitting regions---the high dust column densities and active star formation in these galaxies currently extends well beyond the existing central bulge, as we discuss further in Section~\ref{sec:discussion_bars}.

\subsection{Stellar colors \& the structure of the ISM}
\label{sec:ISM_structure}

One of the most notorious degeneracies in SED fitting is that between dust attenuation and stellar age, and this has long been a concern for SMG samples in particular \citep[e.g.,][]{Hainline_2011, Michalowski_2014, Dudzeviciute_2020}.  Although SED fitting that requires energy balance with the FIR emission can help by providing additional constraints on the dust attenuation \citep[e.g.,][]{daCunha_2015}, the assumptions behind this energy balance have not been thoroughly tested in the high-redshift universe. In particular, the dust model used in such SED fitting codes is calibrated based on the Milky Way \citep[e.g.,][]{Draine_2007} and may not be applicable beyond the local universe and/or in extreme starbursts. 

With this in mind, the relations seen in Figs.~\ref{fig:color-color} and \ref{fig:MdustvsAv} are notable. In particular, the correlation between NIRCam colors and 870$\mu$m surface brightness seen for the majority of the sources (Fig.~\ref{fig:color-color}) implies that the primary driver behind the red stellar colors in these SMGs is \textit{dust} rather than stellar age. Fig.~\ref{fig:MdustvsAv} further implies that the dust-to-stellar distributions in these SMGs are similar (on the same $\sim$kpc-scales) to those in nearby star-forming galaxies, particularly the LIRG NGC2146, which exhibits similar $A_{\rm V}$ for gas and stars \citep[as also seen in high-redshift SMGs;][Taylor et al.\,in prep.]{Birkin_2022PhDT}. The bulk of the SMG data points fall in a similar regime between the foreground dust screen and mixed media models as the \citet{Kreckel_2013} relation, indicating a similar correction factor to the foreground-screen model is applicable for these $z\sim3$ sources as found at $z\sim0$.
This suggests that the ISM in these galaxies consists of a similar combination of mixed dust and emitting material and foreground screen components (c.f. \citealt{Tomicic_2017} for a higher-resolution study in M31), with the remaining scatter likely due to geometric effects and non-uniformity in the dust distribution on scales beneath our resolution.

At low dust mass surface densities ($<$10$^{6}$ M$_{\odot}$ kpc$^{-2}$), it is noteworthy that the SMG data points remain below the limit set by the dust screen model despite the NIRCam data's sensitivity to higher $A_{\rm V}$ values. This suggests that this simple model---based on Milky Way-like dust---remains valid for these $z\sim3$ sources. Meanwhile, at the highest dust mass surface densities probed by these data ($\gtrsim$10$^{7}$--10$^{8}$ M$_{\odot}$ kpc$^{-2}$), the \citet{Kreckel_2013} relation would predict $A_{\rm V}$ values up to $\sim$100 mag, far in excess of what we measure. However, for these particularly high-density regions, our measured $A_{\rm V}$ values likely correspond to lower limits due to the maximum optical depth to the stellar continuum emission that we can detect with NIRCam. 
Given that the nearby galaxies used to calibrate the \citet{Kreckel_2013} relation did not sample such high dust mass surface densities, this does not affect our conclusions.    

In summary, the ISM structure in these SMGs appears largely similar to local star-forming galaxies on $\sim$kpc scales. This similarity implies that some of the most basic assumptions that go into SED-fitting codes such as, e.g., {\sc Magphys} (i.e., the dust model) are still valid  for these $z\sim3$ galaxies.

\subsection{Rest-frame near-infrared and gas/dust bars}
\label{sec:discussion_bars}

Lastly, we discuss the presence of bars in the sources. 
As discussed in detail in \citet{Hodge_2019}, there are clearly visible bar-like sub-structures in the 870$\mu$m images of some of the targets.
While bars are often associated with secular evolution due to their tendency to arise naturally in dynamically cool disk galaxies \citep[e.g.,][]{Athanassoula_1986}
they can also be triggered by interactions \citep{Barnes_1991, Cox_2008, Hopkins_2009_disk_survival}, and thus are not mutually exclusive with the tidal signatures observed here in a large fraction of the sample.  
These bars are also consistent with the most recent theoretical expectations for gas-rich galaxies at high-redshift: Though previous work on bar formation found that the presence of (inert) gas reduces a stellar bar's lifetime \citep{Villa-Vargas_2010} or weakens the bars that form \citep{Athanassoula_2013}, 
more recent work using hydrodynamic N-body simulations to predict bar formation in gas-rich disks at high redshift 
finds that turbulent gas \textit{accelerates} bar formation, predicting bar-like phenomena even in fully gas-dominated turbulent disks \citep{Bland-Hawthorn_2024}. 
Finally, these bars can also be an important mechanism to systematically drive gas inwards in sources with no current evidence for mergers/interaction, helping to explain the intense dusty star formation (and possibly supermassive black hole growth).

Given the high $A_{\rm V}$ values implied in the center of these sources, it is perhaps not surprising that we see no evidence for strong bars in the rest-frame near-infared images presented here, as the stellar components could easily be hidden behind the high implied dust columns.  
Another possible (but more speculative) explanation comes from the work of \citet{Bland-Hawthorn_2024}, who predict that as the gas fraction increases, the role of any stellar bar becomes less important, with mostly gas bars emerging via radial shear flows in galaxies with the highest gas fractions. Considering the high gas fractions inferred for these sources \citep[e.g.,][]{Swinbank_2014}, it is thus possible that their stellar bars are weaker than those in less gas-dominated sources. In either scenario, these dust-rich structures would necessarily be young. 
Notably, \citet{Bland-Hawthorn_2024} find that the bars formed in the most gas-rich disks collapse to form bulges after $\sim$1.5\,Gyr, which would then imply that the bar-like features observed in these sources are an immediate precursor to further significant bulge growth. 
Ultimately, larger samples of sources covering a range of submillimeter flux densities will be necessary to determine the importance of gas/dust and/or stellar bars in the mass assembly of the general SMG population.

\section{Conclusions}
\label{sec:Conclusions}
We have presented JWST NIRCam imaging of 13 $z\sim3$ submillimeter-selected galaxies that have uniquely deep, high-resolution (0.08$''$--0.16$''$) ALMA 870$\mu$m imaging which previously mapped their dust disks on $\sim$0.5--1\,kpc scales. Our main findings are:

\begin{itemize}
    \item All of the sources are securely detected by the JWST NIRcam imaging, including several galaxies which were previously undetected in deep HST 1.6$\mu$m images. For this, observing at or beyond $\sim$3.5$\mu$m ($\simeq$1$\mu$m rest-frame at $z\sim3$) appears to be crucial, as the morphologies of the sources vary dramatically between F200W and F356W (i.e., between $\sim$0.5$\mu$m and $\sim$1$\mu$m rest-frame).  

    \item With the dust-reddened galaxy centers now more visible, the newly revealed rest-frame near-infrared morphologies show some clear similarities to the 870$\mu$m dust continuum images (specifically the position angles and peaks), demonstrating that the spatial offsets previously reported between the 870$\mu$m and HST morphologies were due to strong differential dust obscuration.  

    \item However, we find no correlation between the axis ratio (i.e., ellipticity) derived from the rest-frame near-infrared and 870$\mu$m dust continuum. As the ellipticity is commonly used to set the inclination in dynamical mass estimates, this lack of correlation shows that these inclinations are in fact very poorly determined and that resulting dynamical mass estimates should be used with extreme caution. 

    \item In some sources (particularly those at the highest redshifts), we see convincing evidence that the galaxies are still affected by significant dust obscuration even in the longest-wavelength (F444W) NIRCam filter. For these particularly dusty/high-redshift sources, MIRI imaging will be crucial to more robustly characterize the stellar distributions.

    \item Due to the high level of central obscuration in the galaxies, we find that the median effective radius of the galaxies systematically decreases with increasing NIRCam wavelength. We nevertheless find that the F444W sizes are still 78$\pm$21\% larger than those measured at 870$\mu$m. Thanks to the unique depth and resolution of the 870$\mu$m images available for these targets, this strengthens the challenge posed to recent hydrodynamical simulations claiming the opposite trend \citep{Cochrane_2019, Popping_2022}. However, given the significant obscuration still present in even the longest-wavelength NIRCam filter in at least some of the sources, 
    this result needs to be tested via the creation of stellar mass maps (Li et al., in prep.). 

    \item The NIRCam imaging reveals clear evidence for mergers/interactions (e.g., tidal tails/plumes) in the majority of the sources, with 54$\pm$23\% visually classified as mergers/interactions, 23$\pm$31\% classified as indeterminate, and only 23$\pm$8\% appearing as undisturbed disks (though we caution that these interpretations may miss late-stage mergers or be affected by episodic gas accretion from cooling halos).  Several of the (apparently tidal) features are also detected in 870$\mu$m emission, indicating that the active star formation in these galaxies currently extends well beyond the existing central bulges. 

    \item We find a clear correlation between redder NIRCam colors and 870$\mu$m surface brightness on $\sim$1\,kpc scales. This is a strong indicator that the SMGs have red NIRCam colors due to dust rather than stellar age. We further show that the relation between dust in absorption and dust in emission for the galaxies studied in this work is broadly similar to the trend seen in nearby star-forming galaxies. This suggests that the ISM structure on $\sim$kpc-scales in these $z\sim3$ galaxies is similar to that in $z\sim0$ sources.   

    \item We find no evidence for strong bars in our targets in the rest-frame near-infrared. This suggests that the elongated bar-like structures seen in the high-resolution 870$\mu$m images are highly dust-obscured and/or gas-rich and thus young, implying they are the immediate precursors to further significant bulge growth.  
\end{itemize}

Taken together, these findings suggest we are witnessing heavily obscured and largely interaction-induced bulge formation events at the centers of these massive star-forming galaxies. 
The present study thus demonstrates the joint power of JWST and ALMA for uncovering the morphologies and likely formation histories 
of high-redshift dusty galaxies. Future work will incorporate the multi-wavelength data into resolved SED fitting to explore the resolved physical properties, stellar masses, and resolved kinematics (Li et al.\,in prep; Westoby et al.\,in prep).
Combined with future large samples of sources covering a range of submillimeter flux densities, such studies will be crucial to shed further light on the dominant mechanisms governing the mass assembly of massive, dusty galaxies at high-redshift and the relation of these sources to the larger high-redshift galaxy population. \vspace{9cm} 


\pagebreak 
\begin{acknowledgments}
We thank Joss Bland-Hawthorn for his feedback on the manuscript. JH and BW acknowledge support from the ERC Consolidator Grant 101088676 (``VOYAJ''). 
EdC and JL acknowledge support from the Australian Research Council (projects DP240100589 and CE170100013). 
AMS and IRS acknowledge STFC (ST/X001075/1). 
CCC acknowledges support from the National Science and Technology Council of Taiwan (111-2112M-001-045-MY3), as well as Academia Sinica through the Career Development Award (AS-CDA-112-M02). 
RD acknowledges support from the INAF GO 2022 grant ``The birth of the giants: JWST sheds light on the build-up of quasars at cosmic dawn'' and by the PRIN MUR ``2022935STW'', RFF M4.C2.1.1, CUP J53D23001570006 and C53D23000950006. 
TRG acknowledge funding from the Cosmic Dawn Center (DAWN), funded by the Danish National Research
Foundation (DNRF) under grant DNRF140.
KK acknowledges support from the Knut and Alice Wallenberg Foundation. 
MR is supported by the NWO Veni project ``Under the lens" (VI.Veni.202.225). 
This work is based in part on observations made with the NASA/ESA/CSA James Webb Space Telescope. The data were obtained from the Mikulski Archive for Space Telescopes at the Space Telescope Science Institute, which is operated by the Association of Universities for Research in Astronomy, Inc., under NASA contract NAS 5-03127 for JWST. These observations are associated with program \#\,2516.
Support for program \#\,2516 was provided by NASA through a grant from the Space Telescope Science Institute, which is operated by the Association of Universities for Research in Astronomy, Inc., under NASA contract NAS 5-03127.
 This paper makes use of the following ALMA data: ADS/JAO.ALMA\#2016.1.00048.S, 
 ADS/JAO.ALMA\#2012.1.00307.S, 
and ADS/ JAO.ALMA\#2011.1.00294.S. 
ALMA is a partnership
of ESO (representing its member states), NSF
(USA) and NINS (Japan), together with NRC (Canada)
and NSC and ASIAA (Taiwan) and KASI (Republic
of Korea), in cooperation with the Republic of Chile.
The Joint ALMA Observatory is operated by ESO,
AUI/NRAO and NAOJ. The Hubble Source Catalog is based on data from the Hubble Legacy Archive, which is a collaboration between the Space Telescope Science Institute (STScI/NASA), the Space Telescope European Coordinating Facility (ST-ECF/ESAC/ESA) and the Canadian Astronomy Data Centre (CADC/NRC/CSA). SK acknowledges advice and support from Anton Koekemoer, Armin Rest, and Pablo Perez Gonzalez for the reduction of NIRCam data. 
This research made use of {\sc astropy}, a community developed core Python package for astronomy (\citealt{astropy:2013, astropy:2018}) hosted at http://www.astropy.org/, matplotlib (\citealt{Hunter_2007}), {\sc numpy} (\citealt{Walt_2011}), {\sc scipy} (\citealt{Jones_2001}), and of {\sc topcat} (\citealt{Taylor_2005}).
\end{acknowledgments}

%

\vspace{5mm}





\appendix

\section{Additional Tables \& Figures}
\label{sec:additionalfigs}

\begin{deluxetable*}{ccccc}
	\tablecolumns{5}
	\tabletypesize{\footnotesize}
	\tablecaption{NIRCam final astrometric alignment accuracy}
	\tablehead{
        \colhead{Pointing} &
        \colhead{No. GAIA sources} &
        \colhead{Relative accuracy (mas)\tablenotemark{a}} &
        \colhead{Absolute accuracy (mas)} &
        \colhead{Catalog}}
	\startdata
        1 & 2 & 11 & 65 & Gaia DR3\\ 
        2 & 8 & 13 & 34 & Gaia DR3\\
        3 & 5 & 10 & 19 & Gaia DR3\\
        4 & 4 & 15 & 20 & Gaia DR3\\
        \enddata
        \tablenotetext{a}{relative accuracy is measured between F200W/F356W/F444W filters.}
	\label{tab:astrometry}
\end{deluxetable*}

Table~\ref{tab:astrometry} shows the astrometric accuracy we estimate for each of the NIRCam pointings. Both relative and absolute accuracy are indicated, as well as the number of Gaia stars (with catalog proper motions) available for the absolute astrometric alignment.

\begin{figure*}[t!]
\centering
\includegraphics[width=0.99\textwidth]{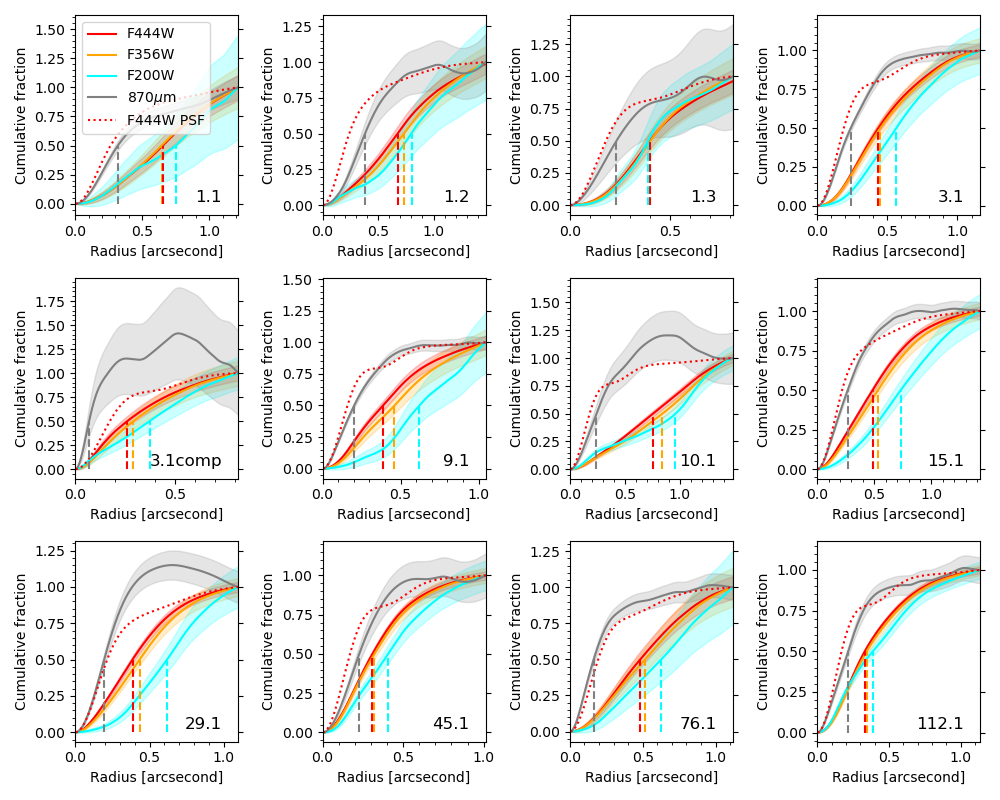}
\caption{Cumulative fraction of integrated flux density in JWST NIRCam and ALMA 870$\mu$m versus radius along the major axis for the targeted SMGs (including 3.1-comp, and excluding 17.1 due to confusion with its optically bright companion). The dotted line indicates the PSF in the F444W filter. The dashed lines indicate the effective radius in each filter, which we find to generally decrease with increasing NIRCam wavelength in these galaxies. Note that the curve corresponding to 870$\mu$m may stray above 1.0 or even decrease with radius due to the presence of noise in the ALMA imaging, indicated by the gray-toned band (representing $\pm$1$\sigma$). In the majority of the sources, the effective radius decreases with increasing NIRCam wavelength, and the effective radius of the ALMA 870$\mu$m emission is more compact than that from even the reddest JWST NIRCam filter. 
}
\label{fig:NClightcurves}
\end{figure*}

Fig.~\ref{fig:NClightcurves} shows the cumulative fraction of integrated flux density versus radius in each of the three NIRCam filters for the targeted SMGs (including 3.1-comp, and excluding 17.1 due to confusion with its optically bright companion). The corresponding curve for the highest-resolution 870$\mu$m image available for each source is also shown.  
These curves are not deconvolved from their (respective) PSFs, but this has a minor effect given that the sources are well-resolved. A curve corresponding to the F444W PSF, which has the lowest resolution of the NIRCam filters, is shown as an example. 

\bibliography{ref}{}
\bibliographystyle{aasjournal}



\end{document}